\documentclass[aps,prx,showpacs,twocolumn,superscriptaddress,nobibnotes]{revtex4-2}
\UseRawInputEncoding
\RequirePackage{subfigure}
\usepackage{lipsum}
\usepackage{graphicx}
\usepackage{color}
\usepackage{enumerate}
\usepackage{amsmath}
\usepackage{bm}
\usepackage{diagbox}
\usepackage{amssymb}
\usepackage{stmaryrd}
\usepackage{multirow}
\usepackage{bbm}
\usepackage[normalem]{ulem}
\usepackage{float}
\usepackage{longtable,booktabs}
\usepackage[dvipsnames,table,xcdraw]{xcolor}
\usepackage[normalem]{ulem}
\usepackage{physics}
\usepackage{url}
\usepackage{soul} 
\usepackage[colorinlistoftodos]{todonotes}
\usepackage{verbatim}
\usepackage{graphicx}
\usepackage{subfigure}
\graphicspath{{figures/}}
\usepackage{appendix}
\usepackage{siunitx}
\definecolor{darkblue}{rgb}{0.1,0.2,0.6} \definecolor{darkred}{rgb}{0.8,0.1,0.2}
\usepackage[colorlinks,citecolor=darkblue,linkcolor=darkred,urlcolor=darkblue]{hyperref}
\usepackage[all]{hypcap} 
\usepackage[draft]{pdfcomment}

\newcommand{\psir}{\ket{\psi}}

\newcommand{\px}{\widetilde{\bm{x}}}

\newcommand{\wx}{\widetilde{x}}

\begin{document}

\title{Gauge Invariant and Anyonic Symmetric Transformer and RNN Quantum States for Quantum Lattice Models}
\author{Di Luo}
\thanks{Co-first authors.}
\affiliation{Department of Physics,  University of Illinois at Urbana-Champaign, IL 61801, USA} 
\affiliation{IQUIST and Institute for Condensed Matter Theory and NCSA Center for Artificial Intelligence Innovation,  University of Illinois at Urbana-Champaign, IL 61801, USA}
\affiliation{The NSF AI Institute for Artificial Intelligence and Fundamental Interactions}
\affiliation{Center for Theoretical Physics, Massachusetts Institute of Technology, MA, 02139, USA}
\author{Zhuo Chen}
\thanks{Co-first authors.}
\affiliation{Department of Physics, University of Illinois at Urbana-Champaign, IL 61801, USA}
\affiliation{The NSF AI Institute for Artificial Intelligence and Fundamental Interactions}
\affiliation{Center for Theoretical Physics, Massachusetts Institute of Technology, MA, 02139, USA}

\author{Kaiwen Hu}
\affiliation{Department of Physics, University of Illinois at Urbana-Champaign, IL 61801, USA}
\affiliation{Department of Physics, University of Michigan, Ann Arbor, Michigan 48109, USA}
\affiliation{Department of Mathematics, University of Illinois at Urbana-Champaign, Urbana, IL 61801, USA}

\author{Zhizhen Zhao}
\affiliation{Department of Electrical and Computer Engineering and CSL, University of Illinois at Urbana-Champaign, Urbana, IL 61801, USA}

\author{Vera Mikyoung Hur}
\affiliation{Department of Mathematics, University of Illinois at Urbana-Champaign, Urbana, IL 61801, USA}

\author{Bryan K. Clark}
\affiliation{Department of Physics, University of Illinois at Urbana-Champaign, IL 61801, USA}
\affiliation{IQUIST and Institute for Condensed Matter Theory and NCSA Center for Artificial Intelligence Innovation,  University of Illinois at Urbana-Champaign, IL 61801, USA}

\begin{abstract}
Symmetries such as gauge invariance and anyonic symmetry play a crucial role in quantum many-body physics. We develop a general approach to constructing gauge invariant or anyonic symmetric autoregressive neural network quantum states, including a wide range of architectures such as Transformer and recurrent neural network (RNN), for quantum lattice models.  These networks can be efficiently sampled and explicitly obey gauge symmetries or anyonic constraint. We prove that our methods can provide exact representation for the ground and excited states of the 2D and 3D toric codes, and the X-cube fracton model. We variationally optimize our symmetry incorporated autoregressive neural networks for ground states as well as real-time dynamics for a variety of models.  We simulate the dynamics and the ground states of the quantum link model of $\text{U(1)}$ lattice gauge theory, obtain the phase diagram for the 2D $\mathbb{Z}_2$ gauge theory, determine the phase transition and the central charge of the $\text{SU(2)}_3$ anyonic chain, and also compute the ground state energy of the SU(2) invariant Heisenberg spin chain. Our approach provides powerful tools for exploring condensed matter physics, high energy physics and quantum information science.  
\end{abstract}

\maketitle

\section{Introduction}

In recent years, there has been a growing interest in machine learning approaches to simulating quantum many-body systems~\cite{Carleo602,Han_2020,Choo_2019,rnn_wavefunction,Luo_2019,paulinet,PhysRevResearch.2.033429,quantum_circuit,gutierrez2020real,topo_wf,Gao2017,Glasser_2018,Vieijra_2020,Nomura_2017,Schmitt_2020,Stokes_2020,Vicentini_2019,Torlai2018,PhysRevE.101.023304,PhysRevLett.126.032001,PhysRevB.99.214306,PhysRevLett.122.250502,PhysRevLett.122.250501}. An important step in this direction is the use of neural networks, e.g. restricted Boltzmann machines, to represent variational wave functions. However, many neural networks do not automatically enforce the symmetries of physical models.  A considerable amount of work has been devoted to remedy the deficiency for several classes of global symmetries, such as translational symmetry~\cite{Choo_2019}, discrete rotational symmetry~\cite{Choo_2019}, global $\text{U}(1)$ symmetry~\cite{rnn_wavefunction}, and anti-symmetry~\cite{Luo_2019,paulinet,PhysRevResearch.2.033429}.

In addition to global symmetries, local symmetries can be encoded through gauge invariance. The notion of gauge invariance is crucial in quantum mechanics. In high energy physics, theory is required to be invariant under the action of gauge symmetry groups~\cite{kogut1975hamiltonian}. Gauge invariance appears naturally in various condensed matter physics models. For example, topological states of
toric code and double semion models arise as the ground states of their gauge-invariant Hamiltonians~\cite{Levin_2005,Kitaev_2003}.  Also, novel quantum matter such as fracton is the ground state of a Hamiltonian where the subsystem symmetry is gauged~\cite{Shirley_2019}. In quantum information, various quantum error correction codes can be viewed as eigenstates in a certain gauge-invariant code space~\cite{Cui_2020}. Besides gauge symmetries, anyonic symmetry is another important local constraint that arises in exotic phases of matter~\cite{Feiguin_2007,Trebst_2008} and topological quantum computation~\cite{field2018introduction}. The study of quantum lattice models with gauge or anyonic symmetries is significant to enhance our understanding of high energy physics, condensed matter physics, and quantum information science.

Simulating quantum many-body gauge theory is exponentially costly. There has been much effort to efficiently simulate quantum lattice gauge theory with both digital and analog quantum computers~\cite{Ba_uls_2020}, but more effort is required experimentally to achieve good fidelity.  Two standard approaches to simulating gauge theory classically are stochastic, integrating an effective Lagrangian by sampling, and variational. 
When simulating gauge theory, the stochastic approach naturally obeys gauge invariance but is plagued with exponential costs associated with the sign problem in models with finite density of fermions or involving quantum dynamics~\cite{rebbi_1983}.  The variational approach overcomes the difficulty by being constrained to an approximate variational space. Imposing gauge symmetries in the variational approach is particularly important and challenging as, otherwise, lower energy states can exist in the gauge-violating part of a Hilbert space.  Therefore gauge symmetries must be explicitly constrained.  While the stochastic approach has been well studied, there have been limited attempts at using the variational approach for gauge theory. Tensor networks can be readily applied to gauge theory in one dimension and ongoing efforts are required to work with challenges in higher dimensions~\cite{Ba_uls_2020,magnifico2020lattice}. A variational approach based on gauge equivariant networks has been introduced very recently~\cite{luo2020gauge,Kanwar_2020,boyda2020sampling}.  

We develop for the first time, a general approach to constructing gauge invariant or anyonic symmetric (such as the fusion rule for anyons) autoregressive neural networks (AR-NN) for quantum lattice models.
Autoregressive neural networks, such as recurrent neural networks (RNN)~\cite{gru,lstm}, pixel convolutional neural networks (pixelCNN)~\cite{pixelcnn}, and Transformers~\cite{transformer},  have revolutionized the fields of computer vision and language translation and generation, among many others.  Autoregressive neural networks quantum states have recently been introduced in quantum many-body physics~\cite{luo2020autoregressive, rnn_wavefunction, autoregressive_variational} and shown to be capable of representing volume law states (as one generically needs in dynamics) with a number of parameters that scale sub-linearly ~\cite{Levine_2019}. A central feature of AR-NN is their capability of exactly sampling configurations from them.  This is to be contrasted with the standard approach of sampling configurations by doing a random walk over a Markov chain, which is often plagued with long equilibration times and non-ergodic behaviors. We construct gauge invariant AR-NN for the quantum link model of U(1) lattice gauge theory~\cite{PhysRevD.11.395}, $\mathbb{Z}_N$ gauge theory, and anyonic symmetric AR-NN for $\text{SU}(2)_k$ anyons.  We demonstrate the exact representation of gauge invariant AR-NN for the ground and excited states of the 2D~\cite{Kitaev_2003} and 3D~\cite{Hamma_2005} toric codes, and the X-cube fracton model~\cite{Vijay_2016}. We optimize our symmetry incorporated AR-NN for 
the quantum link model, the 2D toric code in a transverse field, the 1D Heisenberg chain with SU(2) symmetry, and the $\text{SU}(2)_3$ anyonic chain~\cite{Feiguin_2007,Trebst_2008,Gils_2009}, to obtain ground states accurately and extract phase diagrams and various dynamic properties.

\section{Construction for Gauge or Anyonic Symmetries}
\begin{figure*}[ht]
    \centering
    \includegraphics[scale=0.5]{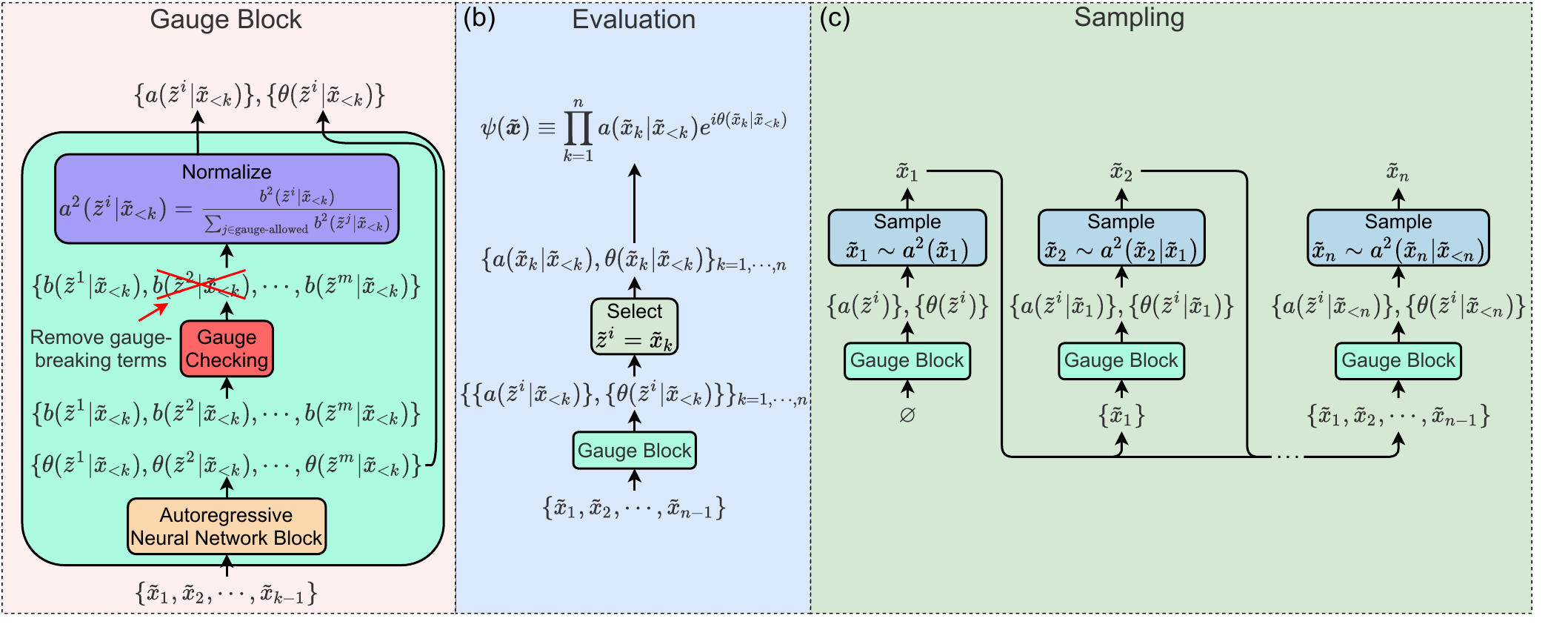}
    \caption{Autoregressive parameterization of wave function with $n$ composite particles. (a) Gauge block. The input $\left\{\widetilde{x}_1, \widetilde{x}_1, \dots, \widetilde{x}_{k-1}\right\}$ is processed through the autoregressive neural network block (see Appendix~\ref{app:nn} for details), to output amplitude and phase parts.  
    The amplitude part goes through gauge checking, which removes the gauge breaking terms. Afterwards, the square of the amplitude is normalized. (b) Evaluation process. The evaluation process can be performed in parallel for all the input sites. Given the input $\{\widetilde{x}_k\}$, the gauge block simultaneously generates amplitudes and phases for all sites. We then select the correct amplitudes and phases based on the input configuration for each site and construct the wave function from the selected amplitudes and phases. (c) Sampling process. The sampling is done sequentially for each site. We begin with no input and generate the amplitude and phase for the first site. The configuration of the first site is then sampled from the square of the amplitude. Afterwards, we feed the first sample into the gauge block to obtain the second sample. This process continues until we obtain the whole configuration.}
    \label{fig:auto_para}
\end{figure*}

Our goal in this work is to generate autoregressive neural networks (AR-NN) which variationally represent wave functions of quantum lattice models and explicitly obey their gauge symmetries---i.e. given a set of gauge symmetry operators $\{G_i\}$ with local support, we would like to construct a wave function $\psir$ such that $G_i \psir = \psir$ for each $i$.  To do this, we will work within the `gauge basis' $\{\ket{x} \}$ which is diagonal in the gauge,    $ \ip{x}{\psi} = \mel{x}{G_i}{\psi}$.  A sufficient condition of gauge invariance of the wave function is to ensure  that the gauge-violating basis elements $\ket{x}$ have zero amplitude in $\psir$.  Throughout this work, we will primarily work with gauges $G_i$ which are local---i.e. $G_i\ket{x} $ only affects a compact range of sites within the vicinity of site $i$.

While we would typically want our AR-NN to take as input the configuration $\{x_1,x_2,\dots,x_n\}$  and evaluate  $\psi(x_1,x_2,\dots,x_n)$, we will find it useful to instead evaluate $\psi(\px)$
where  $\px \equiv \left\{\wx_1, \wx_2, \dots, \wx_{n}\right\}$, $\wx_i \equiv  (x_{i_1},x_{i_2},\dots,x_{i_v})$, is a composite particle specifying the configuration of not only site $i$ but also some number of nearby sites. The motivation for working with composite particles is that a particular local gauge constraint $G_i$ might only depend on composite particle $\wx_i$ (and potentially $\wx_{i+1}$), making it easier to apply the gauge constraints.  Different composite particles can naturally overlap in physical sites and we will simply augment our gauge constraints to require that the configurations of the composite particles agree on the state of a physical site---i.e. basis states of composite particles which map to disagreeing physical states should also have zero amplitude.  

AR-NN perform two functions: sampling and evaluation.   AR-NN can sample configurations $\widetilde{\bm{x}}$ from $|\psi(\widetilde{\bm{x}})|^2$.  This is done sequentially (in some pre-determined order) one composite particle $\widetilde{x}_i$ at a time; the probability to sample $\widetilde{x}_i$ is equal to 
$a^2(\widetilde{x}_i| \wx_{<i})$ where $a(\widetilde{x}_i| \wx_{<i})$ is a function which returns the conditional amplitude. Evaluation of the AR-NN gives a value $\psi(\widetilde{\bm{x}}) = \prod_{i=1}^n a(\widetilde{x}_i| \wx_{<i}) e^{i\theta(\widetilde{x}_i| \wx_{<i})}$ where 
$\theta(\widetilde{x}_i| \wx_{<i})$ is a function which returns the conditional phase. 
Both evaluation and sampling rely on the existence of a gauge block which takes  $\widetilde{x}_1,\dots,\widetilde{x}_{k-1}$ and outputs the possible values $\{\widetilde{z}^i\}$ of $\widetilde{x}_k$ along with their respective amplitudes $a(\widetilde{z}^i|\widetilde{x}_{<k})$ and phases  $\theta(\widetilde{z}^i|\widetilde{x}_{<k})$, ensuring that the amplitude of any configuration which is going to violate the gauge constraint is set to zero. 
To build this gauge block, we start with an autoregressive neural network block which returns a list of amplitudes which do not constrain the gauge (such blocks are standard in autoregressive models such as Transformers and RNN); we  then zero out those partial configurations which break the gauge (on the already established composite particles) and renormalize the probabilities in this list  (see Fig.~\ref{fig:auto_para}(a)).  Given the gauge block it is then straightforward to both sample and evaluate  (see Fig.~\ref{fig:auto_para}(b, c)).  Note the probability induced by our AR-NN is different from the probability induced by the AR-NN with only the autoregressive neural network block even if one projects out the gauge-breaking configurations from the latter network.

It is worth noticing that the construction is not limited to gauge theory, but can be generalized to wave functions with either local or global constraints which are checked in the same way as gauge constraints are checked. This will be helpful for describing constraints from certain global symmetries or special algebraic structure, such as the $\text{SU}(2)$ symmetry for the Heisenberg model and the $\text{SU(2)}_k$ fusion rules for non-abelian anyons.

\section{Optimization Algorithms}\label{sec:optimization}
We use AR-NN to calculate both ground states and real-time dynamic properties.  In both cases, we need to optimize our AR-NN.  For ground states, an AR-NN is optimized with respect to energy and for real-time dynamics, we optimize an AR-NN at time-step $t+2\tau$ given a network at time $t$.  We describe the details of these optimizations.  As these optimization approaches are general, we use $x$ to denote a configuration, but for the context of the paper, $x$ should be viewed as a composite particle configuration. 

For the ground state optimization, we stochastically minimize the expectation of energy for a Hamiltonian $H$ and a wave function $\ket{\psi_\theta}$ as
\begin{equation} \label{eq:local_energy}
    \ev{H}{\psi_\theta} \approx \frac{1}{N}\sum_{x \sim \abs{\psi_\theta}^2}^N \frac{H\psi_\theta(x)}{\psi_\theta(x)} \equiv \frac{1}{N}\sum_{x \sim \abs{\psi_\theta}^2}^N E_\text{loc}(x),
\end{equation}
where $N$ is the batch size and the gradient is given by
\begin{equation}
    \pdv{\theta}\ev{H}{\psi_\theta} \approx \frac{2}{N}\sum_{x \sim \abs{\psi_\theta}^2}^N \real\left\{E_\text{loc}(x) \pdv{\theta} \log \psi_\theta^*(x)\right\}.
    \label{eq:variational_gradient}
\end{equation}
We further control the sampling variance~\cite{variance_reduction} by subtracting from $E_\text{loc}(x)$ the average over the batch, $E_\text{avg} \equiv 1/N \sum_{ x \in\textrm{batch}} E_\text{loc}(x)$,  and define the stochastic variance reduced loss function as
\begin{equation}
    \mathcal{L}_g = \frac{2}{N}\sum_{x \sim \abs{\psi_\theta}^2}^N \real\Big\{\big[E_\text{loc}(x) - E_\text{avg}\big]  \log \psi_\theta^*(x)\Big\},
\end{equation}
where the gradient is taken on $\log \psi_\theta^*$ using PyTorch's~\cite{paszke2019pytorch} automatic differentiation. 

With this loss function, we also use transfer learning techniques~\cite{roth2020iterative,Sprague_2020}. We train our neural networks in smaller systems and use 
these parameters as the initial starting points for optimizing for larger systems (see Appendix~\ref{app:nn} for details). 

For the dynamics optimization, we use a stochastic version of the logarithmic forward-backward trapezoid method~\cite{iserles_2008}, which can be viewed as a higher order generalization of IT-SWO~\cite{kochkov2018variational} and the logarithmic version of the loss functions in Refs.~\onlinecite{luo2020autoregressive, gutierrez2020real}. We initialize two copies of the neural network $\psi_{\theta(t)}$ and $\psi_{\theta(t + 2\tau)}$. At each time step, we train $\psi_{\theta(t + 2\tau)}$ to match $\left(1+iH\tau\right)\ket{\psi_{\theta(t+2\tau)}} \equiv \ket{\Psi_\theta}$ and  $\left(1-iH\tau\right)\ket{\psi_{\theta(t)}} \equiv \ket{\Phi}$ by minimizing the negative logarithm of the overlap, $-\log \flatfrac{\left(\braket{\Psi_\theta}{\Phi}\braket{\Phi}{\Psi_\theta}\right)}{\left(\braket{\Psi_\theta}\braket{\Phi}\right)}$. (Since we only take the gradient on $\theta(t+2\tau)$, for simplicity, we write $\theta$ for  $\theta(t+2\tau)$ and neglect $\theta(t)$.) The inner products related to $\theta$ can be evaluated stochastically  as 
\begin{align}
    &\braket{\Psi_\theta}{\Phi} \approx \frac{1}{N}\sum_{x \sim \abs{\psi_{\theta}}^2}^N \frac{\Psi_\theta^*(x)\Phi(x)}{\abs{\psi_{\theta}(x)}^2} \equiv \frac{1}{N}\sum_{x \sim \abs{\psi_{\theta}}^2}^N \alpha(x), \\
    &\braket{\Psi_\theta} \approx \frac{1}{N}\sum_{x \sim \abs{\psi_{\theta}}^2}^N \frac{\abs{\Psi_\theta(x)}^2}{\abs{\psi_{\theta}(x)}^2} \equiv \frac{1}{N}\sum_{x \sim \abs{\psi_{\theta}}^2}^N \beta(x).
\end{align}
The gradient of the negative logarithm of the overlap can be evaluated stochastically as
\begin{equation}
\begin{aligned}
    &\pdv{\theta}\left(-\log \frac{\braket{\Psi_\theta}{\Phi}\braket{\Phi}{\Psi_\theta}}{\braket{\Psi_\theta}{\Psi_\theta}\braket{\Phi}{\Phi}}\right) \\
    &\approx \frac{2}{N}\sum_{x \sim \abs{\psi_{\theta}}^2}^N \real\left\{\left[\frac{\beta(x)}{\beta_\text{avg}} - \frac{\alpha(x)}{\alpha_\text{avg}}\right]\pdv{\theta} \log\Psi_\theta^*(x)\right\},
    \label{eq:dynamics_gradient}
\end{aligned}
\end{equation}
where $\alpha_\text{avg}$ and $\beta_\text{avg}$ are respectively the average values of $\alpha(x)$ and $\beta(x)$ over the batch of samples.
We can then define the loss function as 
\begin{equation} \label{eq:dynamics_loss}
    \mathcal{L}_d
    \approx \frac{2}{N}\sum_{x \sim \abs{\psi_{\theta}}^2}^N \real\left\{\left[\frac{\beta(x)}{\beta_\text{avg}} - \frac{\alpha(x)}{\alpha_\text{avg}}\right]\log\Psi_\theta^*(x)\right\},
\end{equation}
where the gradient is taken on $\log\Psi_\theta^*$ using PyTorch's~\cite{paszke2019pytorch} automatic differentiation.

For both optimizations, $\psi_\theta(x)$ is evaluated as described in Fig.~\ref{fig:auto_para}(a) and $x$ is sampled from $\abs{\psi_\theta}^2$ as described in Fig.~\ref{fig:auto_para}(b). The full derivations of the stochastic gradients for both optimizations are in Appendix~\ref{app:gradient}.

 In addition, we extensively use the transfer learning technique, by training on small system sizes before moving on to large system sizes. The transfer learning technique provides a good initialization for neural networks that are trained on large system sizes. We observe that the transfer learning technique in general significantly reduces the number of iterations needed. The details of usage of this technique are described in the captions of each figures. (See more details in Appendix~\ref{app:nn}).

\section{Applications in \\ Quantum Lattice Models}

\subsection{$\text{U}(1)$ Quantum Link Model}\label{sec:QLM}

The quantum link model (QLM) of $\text{U}(1)$ lattice gauge theory in $1 + 1$ dimensions in the Hamiltonian formulation with staggered fermions~\cite{PhysRevD.11.395} is defined as
\begin{equation}
\begin{aligned}
H_\textrm{QLM} = & - \sum _ {i  } \left[ \psi _ { i } ^ { \dagger } U _ { i , i + 1 } \psi _ { i + 1 } + \psi _ { i + 1 } ^ { \dagger } U _ { i , i + 1 } ^ { \dagger } \psi _ { i } \right] \\
& + m \sum _ { i } ( - 1 ) ^ { i } \psi _ { i } ^ { \dagger } \psi _ { i } + \frac { g ^ { 2 } } { 2 } \sum _ { i } E _ { i , i + 1 } ^ { 2 },
\end{aligned}
\label{eq:QLM}
\end{equation}
where $m$ is the staggered fermion mass, $g$ is the gauge coupling, $i=1,2,\dots $ labels the lattice site, 
$\psi_i$ is the fermion operator,  $U_{i, i+1}$ is the link variable and $E_{i, i+1}$ the electric flux for the $\text{U}(1)$ gauge field on link $(i,i+1)$~\cite{PhysRevD.11.395}. We denote by $\ket{q_i}$ the basis state at site $i$, and by $\ket{e_{i, i+1}}$ the basis at link $(i,i+1)$. Each unit cell is defined to include two sites and two links. The operators $E_{i,i+1}$ and $U_{i,i+1}$ satisfy the following commutation relations: $[E_{i,i+1},U_{i,i+1}]=U_{i,i+1}$, $[E_{i,i+1},U^{\dagger}_{i,i+1}]=-U_{i,i+1}^{\dagger}$ and $[U_{i,i+1},U_{i,i+1}^{\dagger}]=2E_{i,i+1}$. The gauge constraint 
is given by the Gauss's law operator $\widetilde {G}_{i} = \psi_{i} ^ { \dagger } \psi _ { i } - E _ { i , i + 1 } + E _ { i - 1 , i } + \frac { 1 } { 2 } [ ( - 1 ) ^ { i } - 1 ]$ such that the ground state $\psir$ satisfies $\widetilde {G}_{i} \psir = 0$ for each $i$.  The QLM has gained growing interests and been studied in different settings in recent years \cite{Ba_uls_2020,magnifico2020lattice,Huang_2019,karpov2020disorderfree,verdel2020variational}. We focus on the (1+1)D QLM with the $S=1/2$ representation for the link operators $U_{i,i+1}$ and $E_{i,i+1}$. Under the Jordan-Wigner transformation, Eq.~\ref{eq:QLM} becomes~\cite{Luo_2020} 
\begin{equation}
\begin{aligned}
     H 
    = & - \sum _ { i } \left[ S^+_i S^+_{i,i+1}S^-_{i+1} + \text{H.c.} \right] \\
    & + m \sum _ { i } ( - 1 ) ^ { i } (S^3_i + \frac{1}{2})+ \frac { g ^ { 2 } } { 2 } \sum _ { i } \frac{1}{4},
\end{aligned}\label{eq:QLM'}
\end{equation}
where $S^{\pm} \equiv S^1 \pm iS^2$, $S^1, S^2, S^3$ are the Heisenberg matrices, and the Gauss's law operator becomes $G_i=S^3_i-S^3_{i,i+1}+S^3_{i-1,i}+\frac{1}{2}(-1)^i$. For the $S=1/2$ representation, the last term on the right side of Eq.~\ref{eq:QLM'} is constant and, hence, can be discarded. 

\begin{figure}[h]
    \centering
    \includegraphics[scale=0.5]{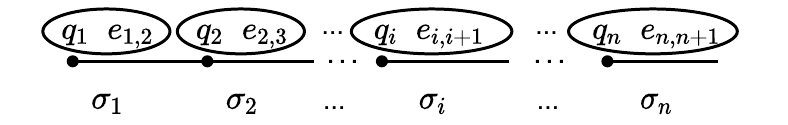}
    \caption{Composite particles for the quantum link model. Each composite particle is defined as $\ket{\sigma_i} \equiv \ket{q_i, e_{i, i+1}}$.  We check Gauss's law between $\ket{\sigma_i}$ and $\ket{\sigma_{i+1}}$. }  
    \label{fig:qlm}
\end{figure}

We define the composite particles of our  gauge invariant AR-NN as in Fig.~\ref{fig:qlm} and choose an order from left to right. Each composite particle $\ket{\sigma_i}$ consists of a fermion $\ket{q_i}$, which can be either $\ket{\bullet}$ or $\ket{\circ}$, and a gauge field in the link $\ket{e_{i, i+1}}$, which can be either $\ket{\rightarrow}$ or $\ket{\leftarrow}$.  Note that in this case the composite particles do not overlap. The Gauss's law operator $G_i$ acts on $\ket{\sigma_i}$ and $\ket{\sigma_{i+1}}$ to determine allowed configurations and so can be checked in the gauge block which generates the composite particle at site $i+1$. 
For example, given $\ket{\sigma_i} = \ket{\bullet\rightarrow}$, $\ket{\sigma_{i+1}}$ can only be $\ket{\bullet\rightarrow}$ or $\ket{\circ\leftarrow}$ if $i$ is even, and $\ket{\circ\rightarrow}$ if $i$ is odd.

\begin{figure}[h]
    \centering
    \includegraphics[width=\linewidth]{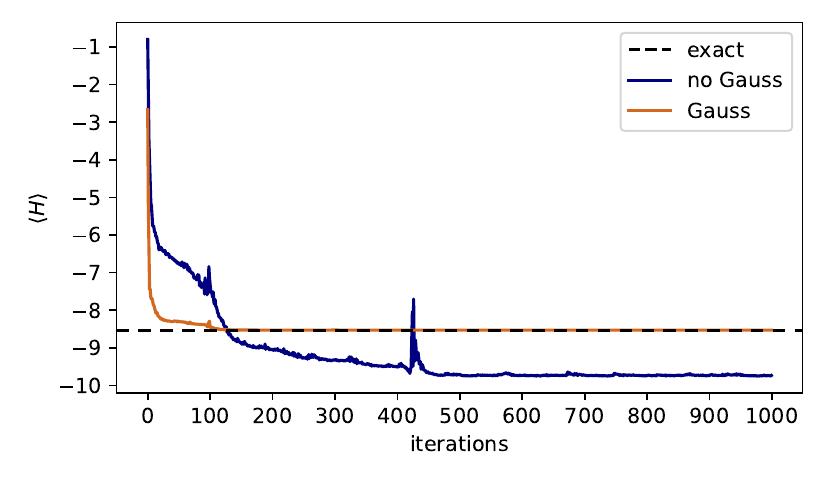}
    \caption{Variational ground state optimization for the 6-unit-cell (12 sites and 12 links) open-boundary QLM for $m=0$ with and without gauge invariant construction. The gauge invariant autoregressive neural network reaches an accurate ground state while the ansatz without gauge constraints arrives at a non-physical state in the optimization. We use the Transformer neural network with 1 layer, 32 hidden dimensions and the real-imaginary parameterization (see Fig.~\ref{fig:parameterization}). The neural network is randomly initialized and is trained for 1000 iterations with 12000 samples in each iteration. The neural network architecture and optimization details are discussed in Appendix~\ref{app:nn}.}  
    \label{fig:qlm_variational}
\end{figure}

\begin{figure}[h]
    \centering
    \includegraphics[width=\linewidth]{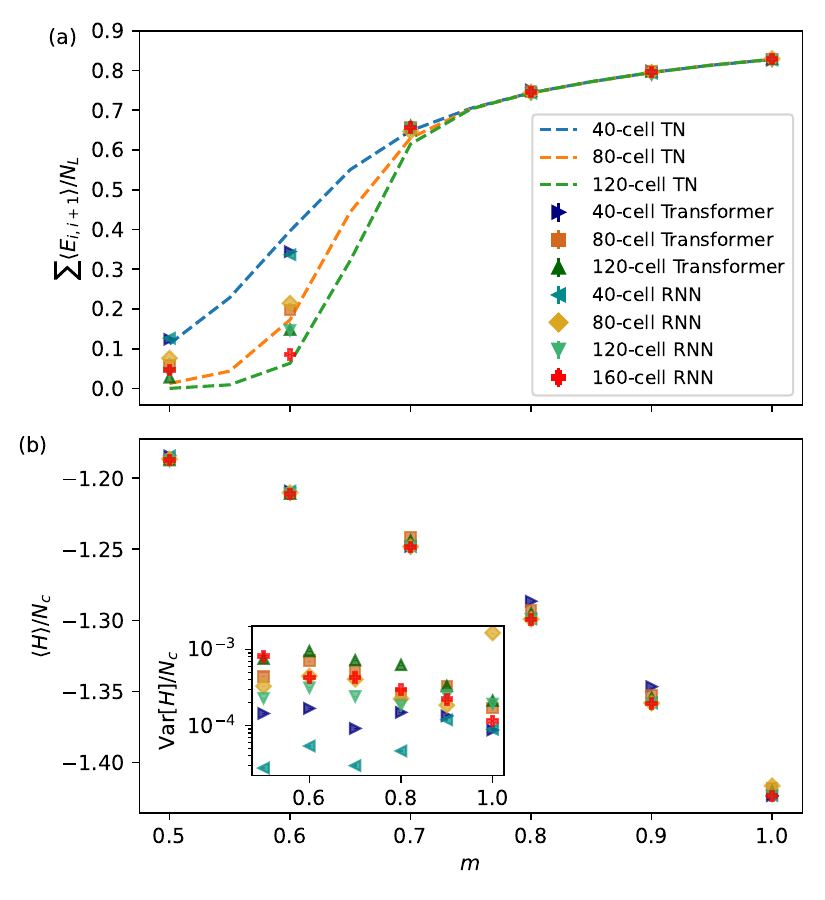}
    \caption{Variational ground state optimization for the open-boundary QLM of different system sizes and different $m$'s with gauge invariant construction. (a) The expectation value of the electric fields averaged over all links, (b) energy and (inset) energy variance per unit cell. We compare our results with the tensor network (TN) results (dashed lines in (a)) from Ref.~\onlinecite{Rico_2014}. The Transformer neural network has 1 layer and 32 hidden dimensions, whereas the RNN has 2 layers and 40 hidden dimensions. For both neural networks, we use the amplitude-phase parameterization  (see Fig.~\ref{fig:parameterization}). The neural networks are randomly initialized. Then they are trained for 3000 iterations with 12000 samples on 40 unit cells. Then, we use transfer learning technique and train the Transformer for 1000 iterations on 80 unit cells and 600 iterations on 120 unit cells. The RNN is then trained for 1000 iterations on 80 and 120 unit cells and 600 iterations on 160 unit cells. The neural network architecture and optimization details are discussed in Appendix~\ref{app:nn}.}
    \label{fig:qlm_variational_fields}
\end{figure}

We implement and variationally optimize this AR-NN for the ground state of Eq.~\ref{eq:QLM'}.  Fig.~\ref{fig:qlm_variational} shows the results for 6 unit cells (i.e. 12 particles) 
which closely match the energy of the exact solution.  
More importantly, the gauge invariant construction guarantees that the solution is in the physical space, while the neural network without gauge constraint (i.e. removing the gauge-checking from the AR-NN) finds a lower energy but non-physical state.

We in addition compute the ground state for 40, 80, 120 and 160 unit cells with both Transformer and RNN (Fig.~\ref{fig:qlm_variational_fields}). The average electric fields are compared with tensor network (TN) results \cite{Rico_2014}. We find that our results (for matching system sizes) are similar to the TN results for both Transformer and RNN.
In addition, we extrapolated the ground state energy for the 160 unit cell model at $m=0.7$ (see Fig.~\ref{fig:qlm_extrapolation} in Appendix.~\ref{sec:additional results}) by linearly extrapolating in variance vs. energy. We find that the extrapolated ground state energy is $-199.7923$, while our lowest energy is $-199.7803\pm0.0005$, giving us a relative error of only $\num{6e-5}$.

\begin{figure}[ht!]
    \centering
    \includegraphics[width=0.95\linewidth]{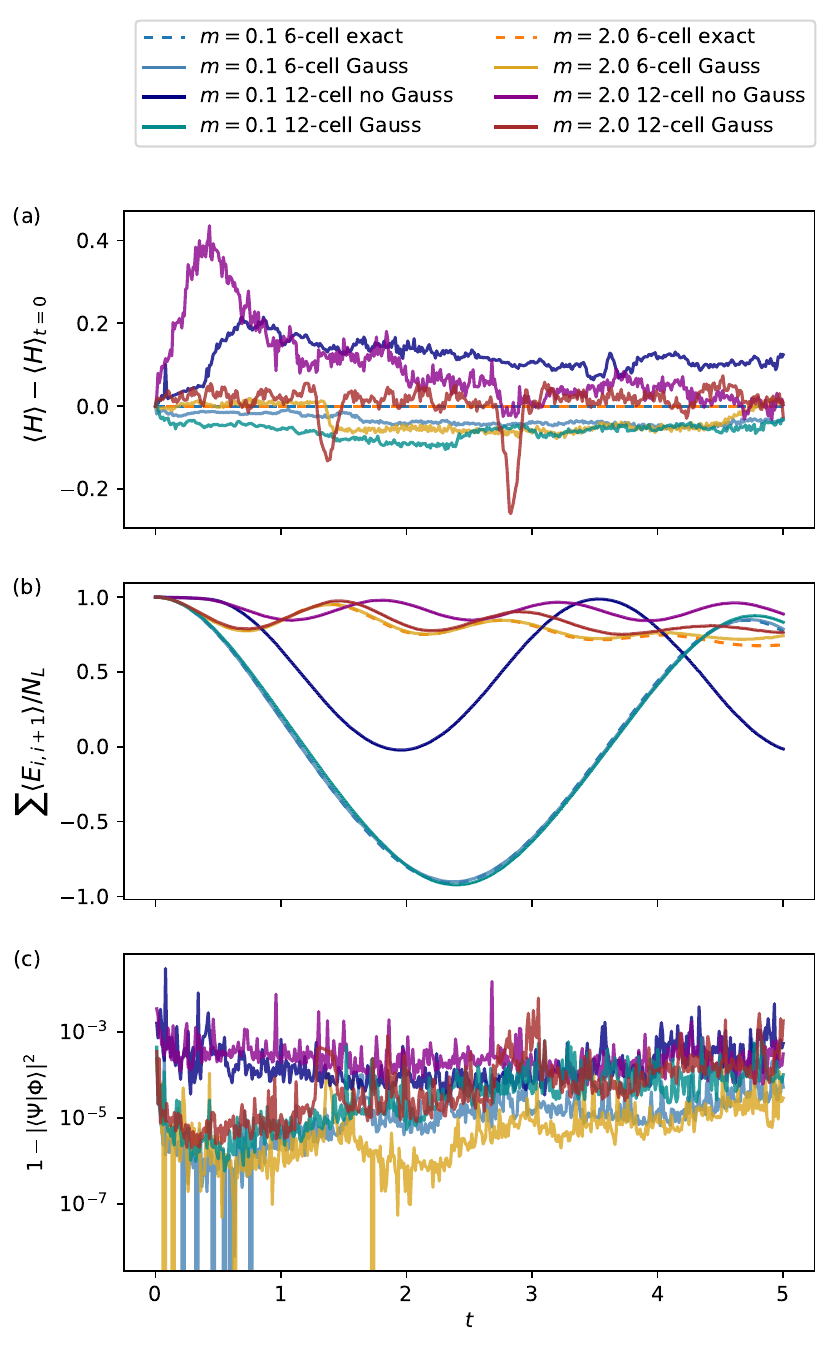}
    \caption{Dynamics for the 6- and 12-unit-cell (12-24 sites and 12-24 links) open-boundary QLM for $m=0.1$ and $m=2.0$ with and without gauge invariant construction. The dashed curves are the exact results from the exact diagonalization for 6 unit cells. (a) The change in the energy during the dynamics. (b) The expectation value of the electric field averaged over all links. (c) The per step infidelity measure, where $\ket{\Psi}$ and $\ket{\Phi}$ are defined in Sec.~\ref{sec:optimization}.
    We use the Transformer neural network with 1 layer, 16 hidden dimensions for 6 unit cells and 32 hidden dimensions for 12 unit cells, and the real-imaginary parameterization (see Fig.~\ref{fig:parameterization}). The initial state is $\ket{\bullet\rightarrow\circ\rightarrow}$ for each unit cell and we train the neural network using the forward-backward trapezoid method with the time step $\tau = 0.005$, 600 iterations in each time step, and 12000 samples in each iteration. The neural network architecture, initialization and optimization details are discussed in Appendix~\ref{app:nn}.}
    \label{fig:qlm_dynamics}
\end{figure}

\begin{figure}[h]
    \centering
    \includegraphics[width=\linewidth]{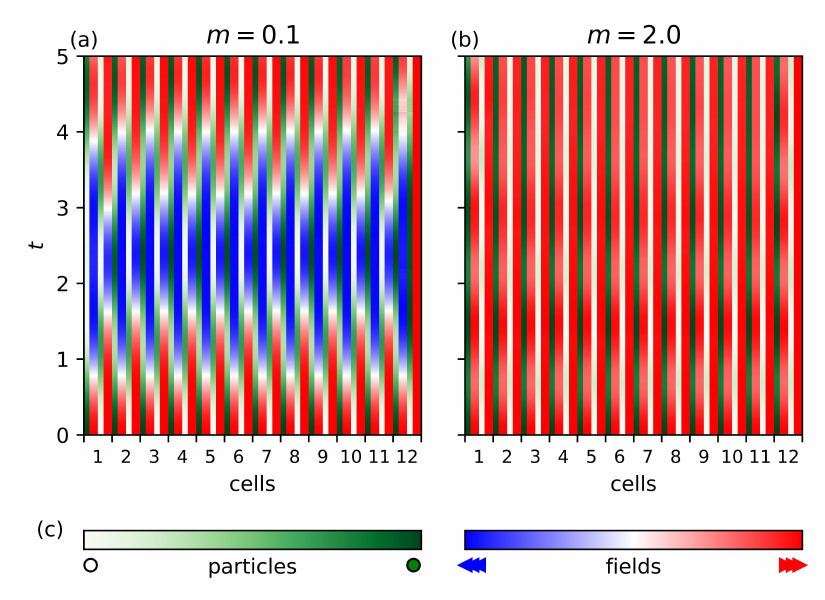}
    \caption{Dynamics of the gauge invariant AR-NN for the 12-unit-cell QLM with (a) $m=0.1$ and (b) $m=2.0$. The ansatz, initialization and optimization are the same as in Fig.~\ref{fig:qlm_dynamics} and are discussed in Appendix~\ref{app:nn}. }
    \label{fig:qlm_color}
\end{figure}

We also consider the real-time dynamics for $m=0.1$ and $m=2.0$ for 6 and 12 unit cells starting with an initial product state with $\ket{\bullet\rightarrow\circ\rightarrow}$ for each unit cell. 
Fig.~\ref{fig:qlm_dynamics}(a) shows the conservation of energy for different ansatzes. The total energy is $-1.2$ for $m=0.1$, and $-24$ for $m=2.0$. We find that our gauge invariant AR-NN captures the correct electric field oscillation and has a lower per step infidelity compared with the non-gauge ansatz (see Fig~\ref{fig:qlm_dynamics}(b) and (c), and, additionally, the anticipated string inversion of the electric flux for small mass (and respectively the static electric flux for large mass) (see Fig~\ref{fig:qlm_color}).

While the current work focuses on the $S=1/2$ representation, our construction can be generalized to an arbitrary $S$ representation. For a higher spin $S$, composite particles can be defined similarly (see Fig.~\ref{fig:qlm}) except that the degree of freedom for each $e_{i,i+1}$ increases to $2S+1$ as $S$ increases.

\subsection{2D $\mathbb{Z}_N$ Gauge Theory}\label{sec:2D toric}

For the 2D toric code~\cite{Kitaev_2003}, consider an $L \times L$ periodic square lattice, where each edge has the basis $\{|0\rangle, |1\rangle\}$. Let $V, P, E$ denote the sets of vertices, plaquettes and edges of the lattice, respectively, such that $|V|=L^2, |P|=L^2, |E|=2L^2$.  
Here we consider the toric code with a transverse magnetic field
\begin{equation}\label{eq:toric_transverse}
    H = H_{TC} - h \sum_{e \in E} \sigma^z_e,
\end{equation}
where $H_{TC}$ is the toric code Hamiltonian
\begin{equation}\label{eq:2d_toric}
    H_{TC} = -\sum_{v\in V} A_v -\sum_{p\in P} B_p,
\end{equation}
$A_v \equiv \prod_{e \ni v} \sigma^z_e$ (the star operator) , $B_p \equiv \prod_{e \ni p} \sigma^x_e$, and $h$ is the strength of the transverse field. 
Note that $A_v$ is the gauge constraint such that the ground state $\psir$ of Eq.~\ref{eq:toric_transverse} and Eq.~\ref{eq:2d_toric} satisfies $A_v \psir = \psir$ for each $v$. 

\begin{figure}[h]
   \includegraphics[scale=0.5]{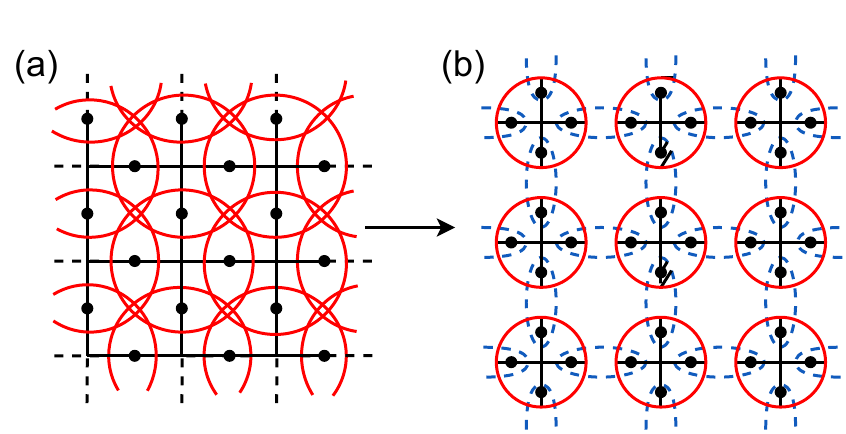}
   \caption{Composite particles for the 2D toric code. (a) Physical structure of 2D toric code with red circles specifying composite particles. Note multiple composite particles share the same physical sites. (b) Composite particles. We define each star as a composite particle (red circle) and check bond consistency for physical sites shared by adjacent composite particles (blue dashed ovals). }
   \label{fig:toric_code_param}
\end{figure} 

The composite particle construction is illustrated in Fig.~\ref{fig:toric_code_param}. We order our consecutive particles by an ``S'' shape going up one row and down the next (see Fig.~\ref{fig:2drnnlayer}(b) in Appendix~\ref{app:nn}).   Two constraints must be checked in the gauge checking process of a gauge block.

When working on the gauge block associated with composite particle $\widetilde{x}_v$, we check that $A_v\ket{\widetilde{x}_v}=\ket{\widetilde{x}_v}$  (despite $A_v$ acting on an entire state, this can be checked locally on a single composite particle).  In addition, composite particles overlap with their four immediately adjacent composite particles. The gauge block for the composite particle at $v$ therefore checks consistency of the physical sites with the first $v-1$ composite particles.  For example, given the configuration $\scriptstyle\ket{\scriptstyle1 \underset{\scriptstyle0}{\overset{\scriptstyle0}{{\color{white}\cdot}}} 1}$ for a composite particle, the composite particle to the right can only be $\scriptstyle\ket{\scriptstyle1 \underset{\scriptstyle0}{\overset{\scriptstyle0}{{\color{white}\cdot}}} 1}$, $\scriptstyle\ket{\scriptstyle1 \underset{\scriptstyle0}{\overset{\scriptstyle1}{{\color{white}\cdot}}} 0}$, $\scriptstyle\ket{\scriptstyle1 \underset{\scriptstyle1}{\overset{\scriptstyle0}{{\color{white}\cdot}}} 0}$ or $\scriptstyle\ket{\scriptstyle1 \underset{\scriptstyle1}{\overset{\scriptstyle1}{{\color{white}\cdot}}} 1}$, as the $\ket{1}$ on the right of the left particle must also be on the left of the right particle. In the ``S'' ordering, there always exists valid choices for each composite particle. For a composite particle that is not the last one, there is an unchosen site which provides freedom of choices to be valid. For the last composite particle, though all sites are fixed, the fixed configuration must be valid because the product of all the Gauss's law constraints is 1 and all previous Gauss's law constraints have been satisfied to be 1. 

\begin{figure}[h]
    \centering
    \includegraphics[width=\linewidth]{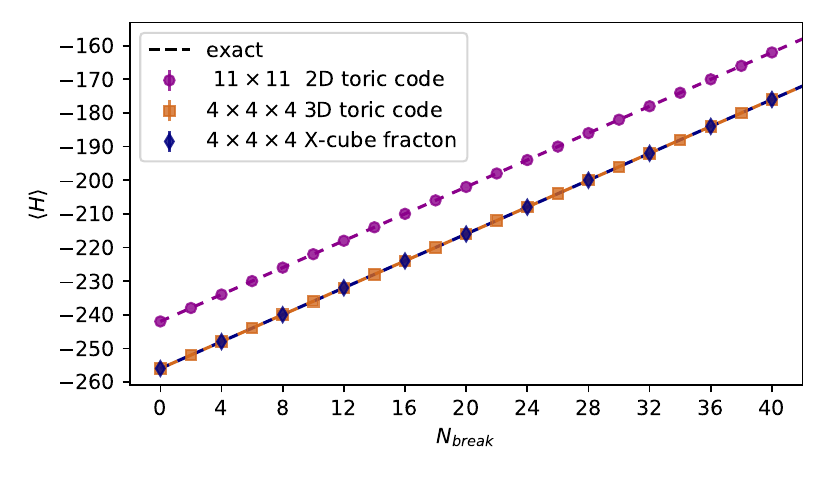}
    \caption{Energies of the analytical constructions of the ground and excited states of the $11\times11$ 2D and $4\times4\times4$ 3D toric code, and the $4\times4\times4$ X-cube fracton model. Here $N_\text{break}$ is the number of Gauss' law violations. The dashed lines are the exact values for each model. The analytical construction generates the same values as the exact up to stochastic errors from sampling. }  
    \label{fig:ana_con}
\end{figure}

We begin by showing that we can analytically generate an AR-NN for the ground state of $H_{TC}$.  One ground state of Eq.~\ref{eq:2d_toric} is $\psir = \prod_{v\in V} (\mathbbm{1}+A_v) \ket{+}^{\otimes n}$ where $\ket{+} = (\ket{0} + \ket{1})/\sqrt{2}$---i.e. an equal superposition of all configurations in the gauge basis which do not violate the gauge constraint.  In our construction, if we use an autoregressive neural network block which gives equal weight to all the configurations  (this is straightforward to arrange by setting the last linear layer's weight matrices to zero and bias vectors to equal amplitudes), we exactly achieve this state.  Checking the $A_v$ does not affect the relative probabilities because it is not conditional involving only one composite particle. On the other hand,
the `gauge constraints' which verify consistency of the underlying state of the sites leave equal probability between all consistent states.  To see this, we examine the effect of the gauge constraint on $|a(\wx_{k}|\wx_{<k})|^2$ for any given $k$, which is the conditional probability of the composite particle $\wx_{k}$. Due to the conditioning from previous composite particles $\{\wx_{<k}\}$,  some sites of the composite particle $\wx_k$ are fixed. 
For the Gauss's law gauge constraints to be $1$, the product of all the unchosen site configurations in $\wx_k$ must be either $1$ or $-1$, depending 
on the chosen site configurations. Let $S_{1} =\{b_{1},\dots,b_{j}|\prod_{r=1}^j b_{r}=1\}$ and $S_{-1} =\{c_{1},\dots,c_{j}|\prod_{r=1}^j c_{r}=-1\}$ be the two possible sets of unchosen site configurations, where $b_r$ and $c_r$ are the configurations of the unchosen sites in $\wx_k$. Consider a function $f: S_{1} \rightarrow S_{-1}$ such that $f(b_{1})=-c_{1}$ and $f(b_{r})=c_{r}$ otherwise. Notice that $f$ is bijective and thus $S_{1}$ and $S_{-1}$ have the same cardinality, implying that after normalization $\abs{a(\wx_{k}|\wx_{<k})}^2$ will have the same amplitude for any $\{\wx_{\leq k}\}$. We can also generate excited states by changing the $A_v$ for a fixed (even) number of vertices to constrain this local eigenvalue to be $-1$ instead of $1$.    We provide a numerical verification of this by computing the energy for an exactly represented tower of ground and excited states in Fig.~\ref{fig:ana_con}.

\begin{figure}[h]
    \centering
    \includegraphics[width=\linewidth]{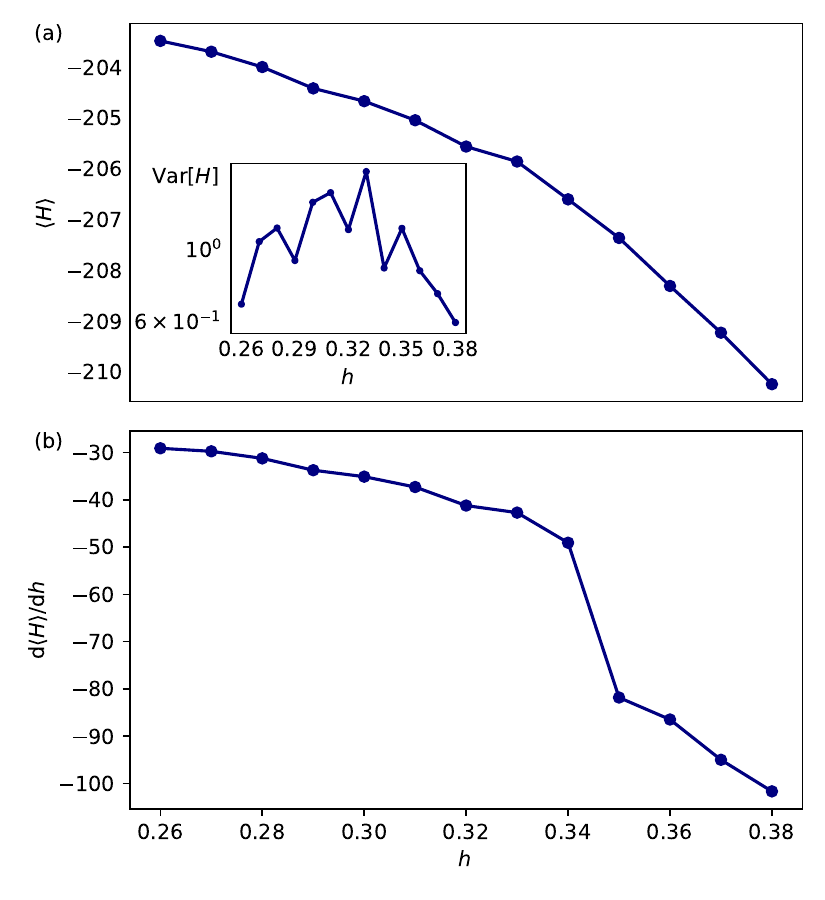}
    \caption{(a) Energy, (inset) energy variance and (b) energy derivative (computed by the Hellman-Feynman theorem\cite{PhysRev.56.340} as $\dv*{\ev{H}}{h} = \ev{\dv*{H}{h}} = -\sum_{e\in E}\ev{ \sigma_e^z}$) versus $h$.   We use the 2D RNN with 3 layers, 32 hidden dimensions and the amplitude-phase parameterization (see Fig.~\ref{fig:parameterization}). We use the transfer learning technique where we first train the neural network on a $6 \times 6$ model for 8000 iterations and then we transfer the neural network to the $10 \times 10$ model for another 1000 iterations. In each iteration, we use 12000 samples. The neural network architecture, initialization and optimization details are discussed in Appendix~\ref{app:nn}.
    }
    \label{fig:toric_code_optimization}
\end{figure}

\begin{figure}[h]
    \centering
    \includegraphics[width=\linewidth]{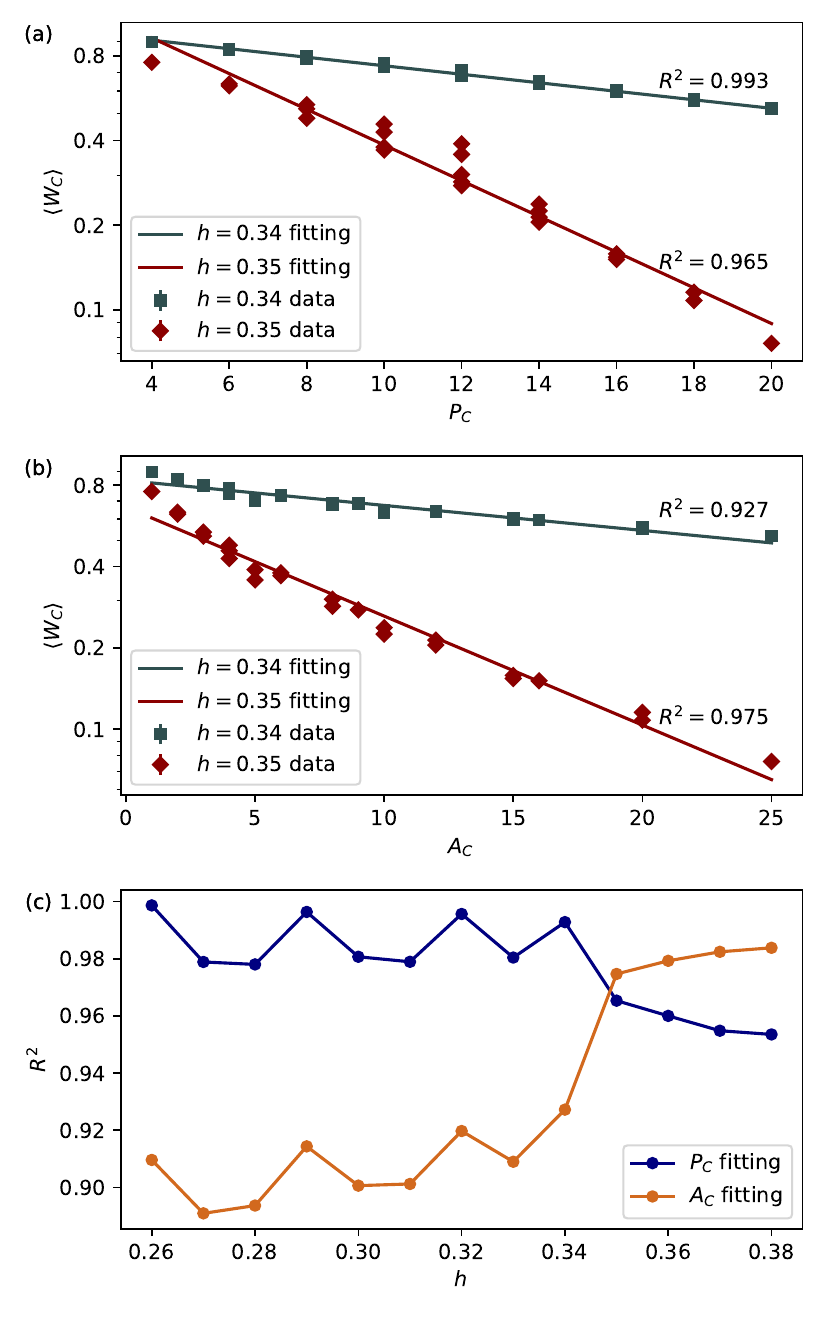}
    \caption{Perimeter and area laws for the $10 \times 10$ 2D toric code. The expectation value of the Wilson loop operator with respect to the (a) perimeter and (b) area of the loop in a log-y scale for $h=0.34$ and  $0.35$.  (c) The fitting of the correlation coefficient $R^2$ for the  perimeter and area laws for different $h$. The ansatz, initialization, and optimization are the same as in Fig.~\ref{fig:toric_code_optimization} and discussed in Appendix~\ref{app:nn}.}  
    \label{fig:toric_code_wilson_loop}
\end{figure}

\begin{figure}[h]
    \centering
    \includegraphics[width=\linewidth]{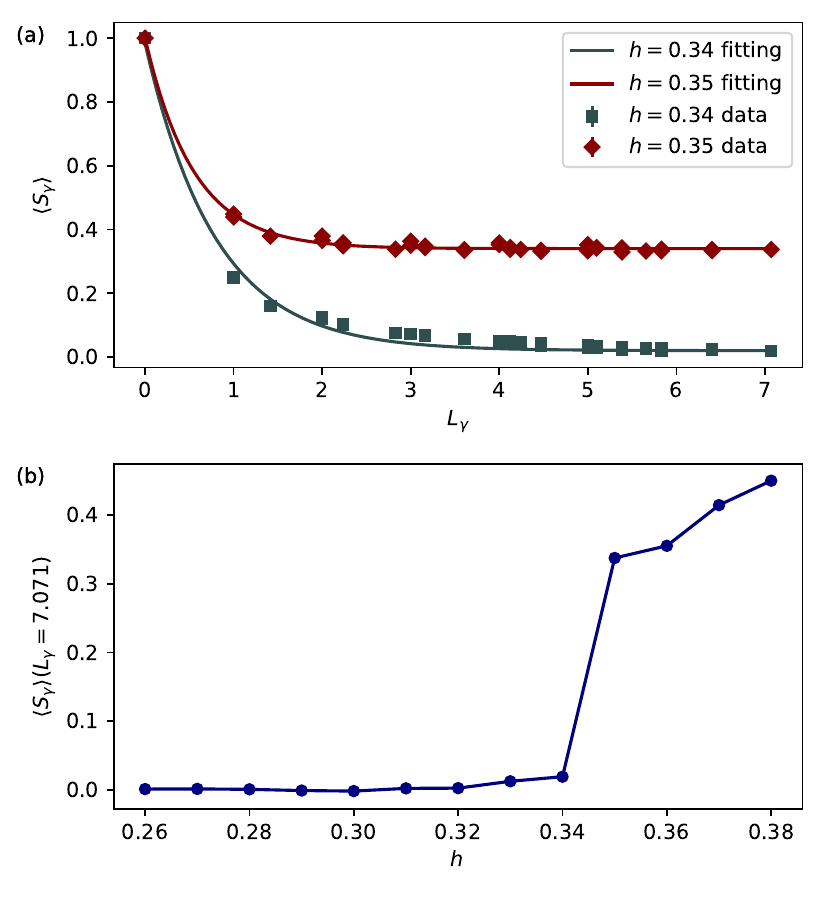}
    \caption{(a) Non-local string correlation function for the $10 \times 10$ 2D toric code between a pair of particle and anti-particle with a distance of $L_y$ apart.  (b) The correlation of a pair of particle and anti-particle at a distance of $L_y=5\sqrt{2}$ for different $h$. The ansatz, initialization, and optimization are the same as in Fig.~\ref{fig:toric_code_optimization} and discussed in Appendix~\ref{app:nn}.}  
    \label{fig:toric_code_correlation}
\end{figure}

\begin{figure}[h]
    \centering
    \includegraphics[width=\linewidth]{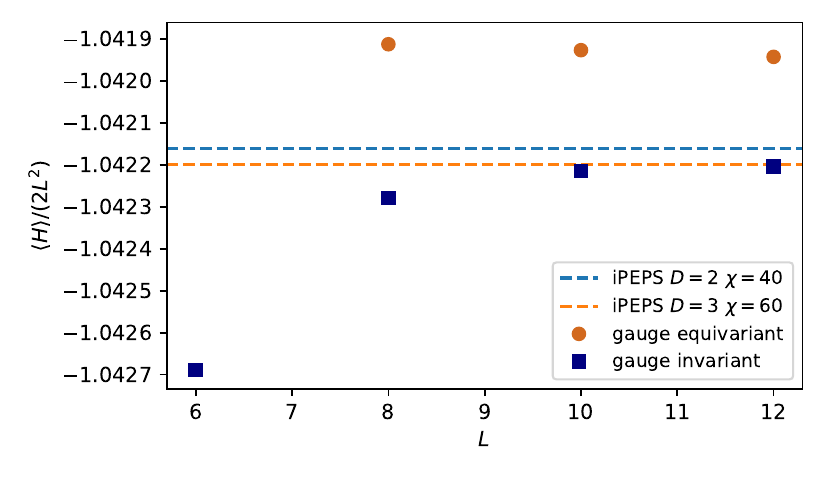}
    \caption{(a) Energy per site and (b) variance of energy per site for $L\times L$ toric code model with $h=0.36$. We compare our results (blue squares) with the iPEPS results of infinite system size (dashed lines) from Ref.~\onlinecite{Crone_2020} and the gauge equivariant neural network results from Ref.~\onlinecite{luo2020gauge}. Notice that due to the difference in the definition of $h$, the $h$ here is twice as large as in Ref.~\onlinecite{Crone_2020}. We use the 2D RNN with 3 layers, 32 hidden dimensions and the amplitude-phase parameterization (see Fig.~\ref{fig:parameterization}). The neural network is randomly initialized and trained for 8000 iterations with 12000 samples on $6\times6$ system. Then, we used the transfer learning technique to train the neural network on $8\times8$ and $10\times10$ systems for another 8000 iterations and on $12\times12$ system for 4000 iterations. The neural network architecture, initialization and optimization details are discussed in Appendix~\ref{app:nn}.
    }
    \label{fig:toric_code_compare}
\end{figure}

With a nonzero value of the external field $h$, the ground state of Eq.~\ref{eq:toric_transverse} is no longer exactly representable, and we variationally optimize our AR-NN to compute the ground state energy.  Fig.~\ref{fig:toric_code_optimization} shows the minimum energy for Eq.~\ref{eq:toric_transverse} for different $h$ and the energy derivative, computed  using the Hellman-Feynman theorem~\cite{PhysRev.56.340}. The toric code is expected to exhibit a quantum phase transition between the topological and trivial phases at an intermediate value of $h$, and the sharp change of the energy derivative around $h=0.34$ is an indicator of this phase transition, which is consistent with the quantum Monte Carlo prediction of $h = 0.328474$ in the thermodynamic limit~\cite{wu2012phase}.
We can additionally  identify the transition by considering the Wilson loop operator $W_C = \prod_{e \in C} \sigma^x_e$ for a closed loop $C$. It is predicted that the topological order phase follows an area law decay, $\ev{W_C} \sim \text{exp}(-\alpha A_C)$, and the trivial phase follows a perimeter law decay, $\ev{W_C} \sim \text{exp}(-\beta P_C)$, where $A_C, P_C$ are the enclosed area and perimeter of the loop $C$~\cite{Gregor_2011}. 
Fig.~\ref{fig:toric_code_wilson_loop} shows the values of $\langle W_c \rangle$ using our variationally optimized AR-NN.  By comparing the respective fits to the area and perimeter laws we again see the transition at $h=0.34$.  Finally, we compare the non-local string correlation operators $S_{\gamma} = \prod_{e \in \gamma} \sigma^z_e$ of our variational states which could be viewed as a measure of the correlation of a pair of excited particle and anti-particle along a path $\gamma$. In the topological order phase the non-local string operators will decay to zero while they will remain constant at the trivial phase~\cite{Zarei_2019}.  In Fig.~\ref{fig:toric_code_correlation}, this is seen clearly on both sides of the transition.  

At $h=0.36$, we additionally benchmark our results with the infinite system size iPEPS results \cite{Crone_2020} and the gauge equivariant results \cite{luo2020gauge} (Fig.~\ref{fig:toric_code_compare}).  Here we use an improved ansatz with $180^\circ$ rotation symmetry defined by $\log \psi_\text{new}(x) = \log (\abs{\psi(x)}^2 + \abs{\psi(R(x))}^2) / 2$, where $R(x)$ rotates configuration $x$ by 180 degrees. We find that our results (at least up to $L=12$) are lower in energy density than the gauge equivariant results and the iPEPS results, which indicates that our approach is very competitive with the state-of-the-art methods.

Our approach can be naturally generalized to 2D $\mathbb{Z}_N$ gauge theory, which can be described in the language of Kitaev's $D(G)$ model with group $G=\mathbb{Z}_N$ (see Appendix~\ref{app: DG_model}). In this case, the basis at each edge becomes a group element in $\mathbb{Z}_N$. Similarly to Fig.~\ref{fig:toric_code_param}, one can define a composite particle over four edges from a vertex and impose gauge invariance. We can also extend our approach to the (1+1)D $\mathbb{Z}_N$ lattice quantum electrodynamics (QED) model, which is discussed in Appendix~\ref{app:zd_qed}.

\subsection{3D Toric Code and Fracton Model}

We turn to gauge invariant AR-NN for the ground and excited states of the 3D toric code~\cite{Hamma_2005} and the fracton model~\cite{Haah_2011,Vijay_2016}. The 3D toric code generalizes the 2D toric code to an $L \times L \times L$ periodic cube where each edge has the basis $\{\ket{0}, \ket{1}\}$. The Hamiltonian takes the same form as the 2D model (see Eq.~\ref{eq:2d_toric}) except that for each $A_v \equiv \prod_{e \ni v} \sigma^z_e$ there are six edges $e$ associated with each vertex $v$. A ground state of the 3D toric cube similarly satisfies $A_v \psir = B_p \psir = \psir$ for each $v, p$. One of the degenerate ground states can also be expressed as $\psir = \prod_{v\in V} (\mathbbm{1}+A_v) \ket{+}^{\otimes n}$. The excited states can be generated by breaking certain constraints from $A_v$ and $B_p$ as in the 2D case. 

\begin{figure}[h]
        \includegraphics[scale=0.5]{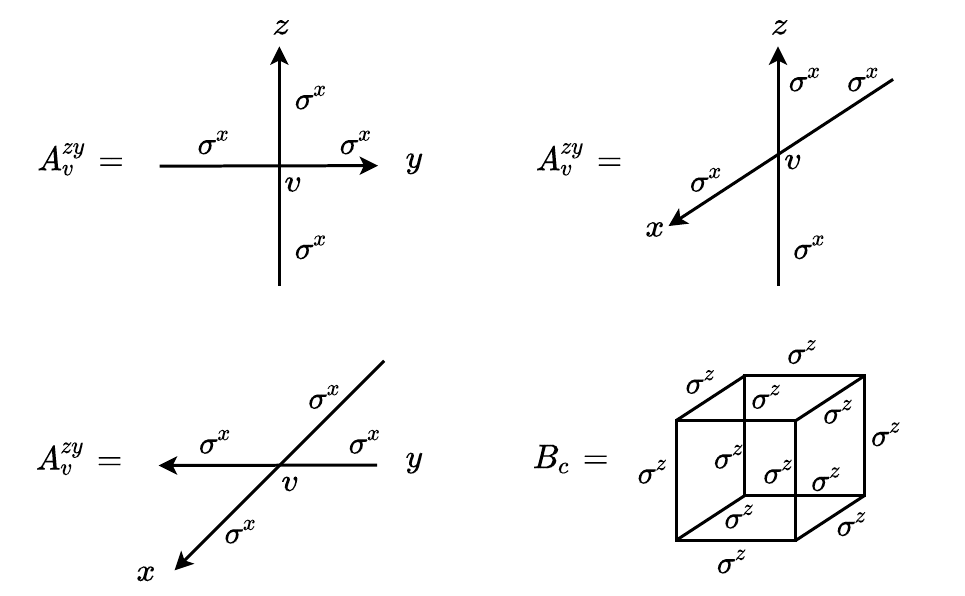}
   \caption{\label{fig:fracton} $A_v^i$ and $B_c$ for the X-cube fracton model.}
\end{figure} 

The X-cube fracton model~\cite{Vijay_2016} is also defined on an $L \times L \times L$ periodic cube where each edge has the basis $\{\ket{0}, \ket{1}\}$. The Hamiltonian takes the following form
\begin{equation}
    H_{\text{fracton}} = -\sum_{v \in V, i} A_v^i -\sum_{c \in C} B_c,
\end{equation}
where $B_c \equiv \prod_{e \in c} \sigma^z_e$ over the edges in a small cube. 
The gauge constraint, i.e. Gauss's law, is $B_c|\psi\rangle = |\psi\rangle$.
There are three $A_v^i \equiv \prod_{e^i \ni v} \sigma^x_{e^i}$ for three choices of $i=zy, xy, xz$, depending on which 2D plane $A_v^i$ acts on. The operators are illustrated in Fig.~\ref{fig:fracton}. A ground state of the X-cube fracton model satisfies $A_v^i \psir = B_c \psir = \psir$ for each $i,v,c$. One of the ground states can be expressed as $\psir = \prod_c (\mathbbm{1}+B_c) |+\rangle^{\otimes n}$. The excited states break some constraints such that $A_v^i \psir = - \psir$ or $B_c \psir = - \psir$ for certain $A_v^i, B_c$.

\begin{figure}[h]
   \includegraphics[scale=0.5]{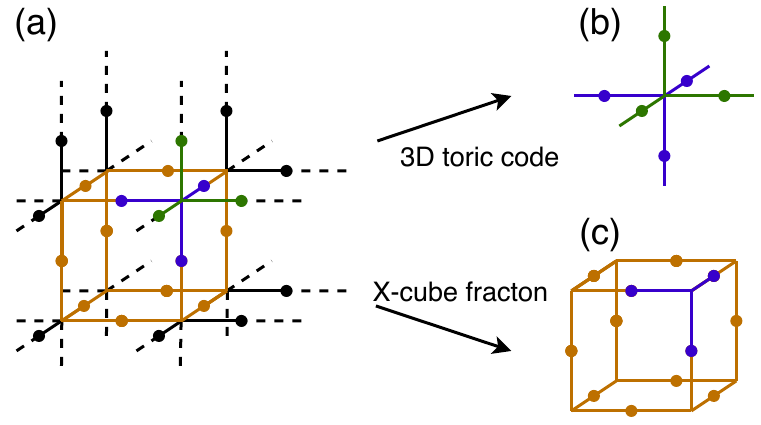}
   \caption{Composite particles of the 3D toric code and X-cube fracton model. The colors are to help identify how composite particles are defined (i.e. which parts of (a) map to (b) and (c)). (a) Physical structure of the 3D toric code and X-cube fracton model. (b) Composite particles of the 3D toric code. We define each star as a composite particle and check bond consistency for adjacent particles similarly to the 2D toric code. (c) Composite particles of the X-cube fracton model. We define each cube as a composite particle and check bond consistency on faces of adjacent particles.}
   \label{fig:3D_param}
\end{figure}

The composite particles 
for the 3D toric code and the X-cube fracton model are illustrated in Fig.~\ref{fig:3D_param}. For the 3D toric code, a composite particle is made up of six particles associated with a vertex. The ground state can be constructed by initializing the bias of the autoregressive neural network to be all the $\ket{+}$ state and imposing gauge checking on each composite particle to be $1$. The excited states can be constructed by forcing even numbers of composite particles to have gauge checking value $-1$. For the X-cube fracton model, a composite particle consists of twelve particles on each mini cube. The ground state comes from initializing all biases of the autoregressive neural network to be $\ket{+}$ and requiring all composite particles to have the gauge checking value $1$. The excited states break the gauge checking value on sets of four nearby composite particles to be $-1$. We numerically verify the exact representations of ground and excited states of the 3D toric code and the X-cube fracton model in Fig.~\ref{fig:ana_con}, where the energy is shown to be exactly the same as the theoretical predictions. 

Our approach can be naturally generalized to the Haah’s code fracton~\cite{Haah_2011} and checkerboard fracton~\cite{Vijay_2016} models. Similarly to 2D $\mathbb{Z}_N$ gauge theory (see Section~\ref{sec:2D toric}), one can consider applying gauge invariant AR-NN to study the 3D $\mathbb{Z}_N$ gauge theory in the context of the 3D toric code or the X-cube fracton model~\cite{Shirley_2019} with an external field.

\subsection{$\text{SU(2)}_k$ Anyonic Chain and $\text{SU(2)}$ Symmetry}

Non-abelian anyons play a crucial role in universal topological quantum computation. Here we consider a chain of Fibonacci anyons, which can be regarded as an $\text{SU(2)}_{k=3}$ deformation of the ordinary quantum spin-1/2 chain~\cite{Gils_2009}. In this model, there is one type of anyon $\tau$ and a trivial vacuum state $\mathbbm{1}$ for each site. The constraint from symmetry requires that $\tau$ and $\mathbbm{1}$ satisfy the following fusion rule: $\tau \otimes \tau = \tau \oplus \mathbbm{1}$, $\tau \otimes \mathbbm{1} = \mathbbm{1} \otimes \tau = \tau$.  
We work directly in this basis where each site is either $\mathbbm{1}$ or $\tau$, generating an anyonic symmetric AR-NN.  We then proceed to work out the entire phase diagram of the Fibonacci anyons. This can be done particularly efficiently compared with standard Monte Carlo sampling~\cite{Vieijra_2020} thanks to the exact sampling of the AR-NN.  

Our anyonic symmetric AR-NN is constructed so that it obeys the anyon fusion rule directly  
by  checking two adjacent input configurations and imposing zero amplitude when both are $\tau$. 
Each anyon is a composite particle and the gauge checking implements the constraint from the anyon fusion rule.

We consider the Hamiltonian~\cite{Trebst_2008}
\begin{equation}\label{eq:ham_anyon}
    H(\theta) = -\cos \theta \sum_i H^{(2)}_i - \sin \theta \sum_i H^{(3)}_i,
\end{equation}
where
the two-anyon interactions can be described by the golden chain Hamiltonian~\cite{Feiguin_2007,Trebst_2008} 
\begin{align}
\begin{split}
    H^{(2)}_i =& \dyad{\mathbbm{1}\tau\mathbbm{1}} + \phi^{-2} \dyad{\tau\mathbbm{1}\tau}\\ &+\phi^{-1}\dyad{\tau\tau\tau}  \\
    &+\phi^{-3/2}\left(\dyad{\tau\mathbbm{1}\tau}{\tau\tau\tau} + \text{H.c.}\right),
\end{split}
\end{align}
and the three-anyon interactions can be described by the Majumdar-Gosh chain Hamiltonian~\cite{Trebst_2008} 
\begin{align}
\begin{split}
    H^{(3)}_i =&\dyad{\mathbbm{1}\tau\tau\mathbbm{1}} +\left(1- \phi^{-2}\right)\dyad{\tau\tau\tau\tau} \\
    &+\left(1 - \phi^{-1}\right)\left(\dyad{\tau\tau\mathbbm{1}\tau} +  \dyad{\tau\mathbbm{1}\tau\tau}\right) \\
    &- \phi^{-5/2}\left(\dyad{\tau\mathbbm{1}\tau\tau}{\tau\tau\tau\tau} + \dyad{\tau\tau\mathbbm{1}\tau}{\tau\tau\tau\tau}  + \text{H.c.} \right) \\
    &+ \phi^{-2}\left(\dyad{\tau\tau\mathbbm{1}\tau}{\tau\mathbbm{1}\tau\tau} + \text{H.c.}\right),
\end{split}
\end{align} 
$\phi = \left(\sqrt{5} + 1\right) / 2$ is the golden ratio. 

\begin{figure}[h]
    \centering
    \includegraphics[width=\linewidth]{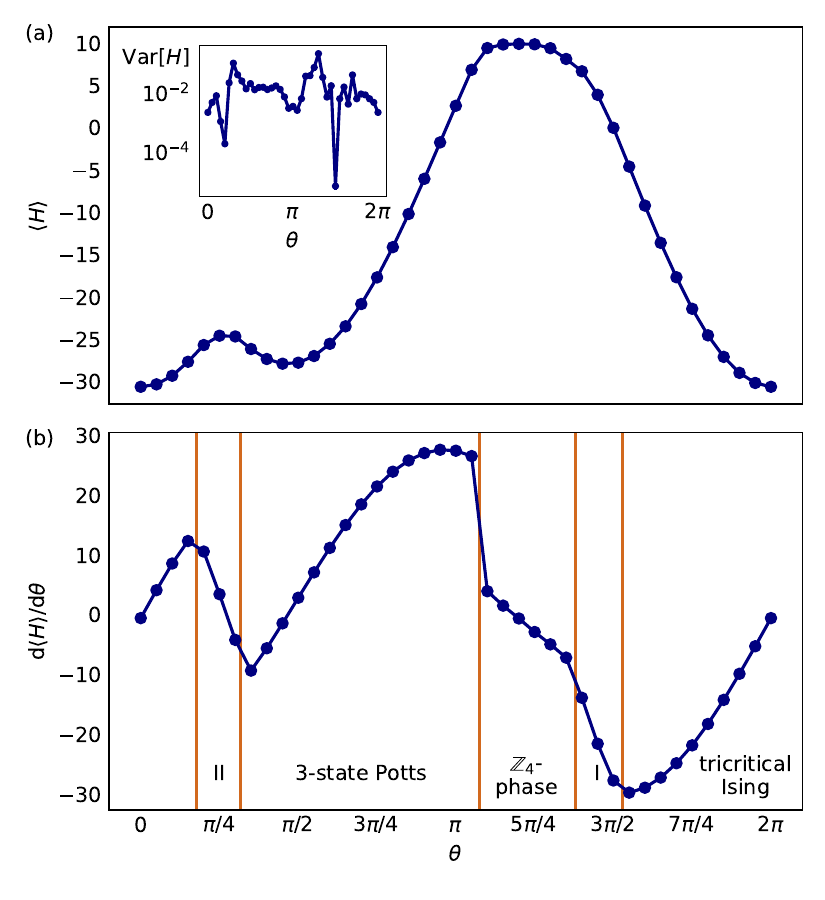}
    \caption{Phase diagram for 40 anyons with the periodic boundary condition. (a) Energy,  (inset) energy variance and (b) energy derivative, computed using the Hellmann–Feynman theorem~\cite{PhysRev.56.340} as $\dv*{\ev{H}}{\theta} = \ev{\dv*{H}{\theta}} = \sin\theta\sum_i\ev*{H_i^{(2)}} - \cos\theta\sum_i\ev*{H_i^{(3)}}$, versus $\theta$. Phase transitions occurs when the energy function is not differentiable. Vertical lines in orange are located at the exact phase transition points~\cite{Trebst_2008}. We use the 1D RNN with 3 layers, 36 hidden dimensions and the real-imaginary parameterization. We use the transfer learning technique where we first train the neural network on a 32-anyon model for 3000 iterations and then we transfer the neural network to the 40-anyon model for another 3000 iterations. In each iteration, we use 12000 samples. The neural network architecture, initialization, and optimization details are discussed in Appendix~\ref{app:nn}.}  
    \label{fig:anyon_phase}
\end{figure}

\begin{figure}[h]
    \centering
    \includegraphics[width=\linewidth]{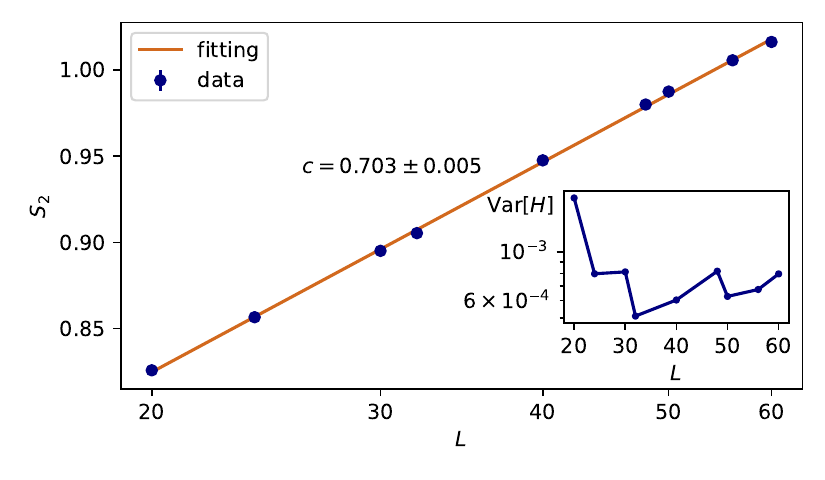}
    \caption{The second Renyi entropy $S_2$ versus the system size $L$ for the optimized AR-NN for the Fibonacci anyons in a golden chain ($\theta=0$) with the periodic boundary condition.  The inset shows the variance of energy of the AR-NN.   The slope of the fitted line is the central charge $c$.  The AR-NN is the 1D RNN with 3 layers, $L$ hidden dimensions, and the amplitude-phase parameterization. The neural network is trained for 8000 iterations with a sample size of 12000. The neural network architecture, initialization, and optimization details are discussed in Appendix~\ref{app:nn}.} 
    \label{fig:central_charge}
\end{figure}

This model is predicted to exhibit five phases with respect to different $\theta$~\cite{Trebst_2008}.  
Fig.~\ref{fig:anyon_phase} shows the optimized energies of the Hamiltonian in Eq.~\ref{eq:ham_anyon} for different $\theta$ and the energy derivative computed using the Hellman-Feynman theorem~\cite{PhysRev.56.340}. The non-differentiable points of the energy derivative indicates the phase transition points, which agree with the conformal field theory prediction. In the special case of $\theta=0$, the model reduces to the  
Fibonacci anyons in a golden chain,
which has a gapless phase~\cite{Trebst_2008}. Using our optimized AR-NN, we compute the second Renyi entropy $S_2$~\cite{rnn_wavefunction}. Since the second Renyi entropy $S_2$ is related to the central charge $c$ under the periodic boundary condition as $S_2 \sim \frac{c}{4} \text{log}(L)$ with system size $L$~\cite{Bazavov_2017}, we then extract the central charge finding a value $c=0.703 \pm 0.005$ very close to the exact result of 0.7 (see Fig.~\ref{fig:central_charge}).

This can be generalized to the ${\text{SU(2)}_k}$ formulation of anyon theory, for which there are $k+1$ species of anyons labeled by $j=0,1/2,1,\dots,k/2$ with the fusion rule of ${\text{SU(2)}_k}$~\cite{Feiguin_2007}. The Hamiltonian 
can be expressed with operators from the representation of the Temperley-Lieb algebra~\cite{Feiguin_2007}.  To construct an anyonic symmetric autoregressive neural network for the general ${\text{SU(2)}_k}$ anyonic chain, one works in the angular momentum basis $\{\ket{\dots, j_{i-1}, j_{i}, j_{i+1},\dots}\}$ where  $j_{i} \in \{0,1/2,1,\dots,k/2 \}$. Since each $j_i$ is included as the $\text{SU(2)}_k$ fusion rule outcome of $j_{i-1}$ and an extra $1/2$ angular momentum, one can view $j_i$ as a composite particle and gauge checking is the fusion rule. 

The Fibonacci anyon is a special case of the $\text{SU(2)}_{k=3}$ formulation, considering the mapping $\tau \mapsto j=1$ and $\mathbbm{1} \mapsto j=0$ and applying $3/2 \times j = 3/2 - j$ from the $\text{SU(2)}_3$ fusion rule to the even-number sites~\cite{Feiguin_2007}. 
Note that this gives a slightly different AR-NN from what is described above. 
Besides the Fibonacci anyon, one can consider the Yang-Lee anyon, which follows the $\text{SU(2)}_3$ fusion rule~\cite{Ardonne_2011}. 

Using this framework, one can also consider the Heisenberg spin chain with SU(2) symmetry since it can be considered as the $\text{SU(2)}_k$ deformation of the ordinary quantum spin-1/2 chain~\cite{Gils_2009} as $k \rightarrow \infty$.  In Appendix~\ref{app:su2}, we provide the detailed construction of an SU(2) invariant autoregressive neural network for the Heisenberg model, which can be viewed as the case of $\text{SU(2)}_{k=\infty}$, and obtain accurate results for the 1D Heisenberg model.

\section{Conclusion}

We have provided a general approach to constructing gauge invariant or anyonic symmetric autoregressive neural network wave functions for various quantum lattice models. These wave functions explicitly satisfy the gauge or algebraic constraints, allow for perfect sampling of configurations, and are capable of explicitly returning the amplitude of a configuration including normalization.  To accomplish this, we have upgraded standard AR-NN in such a way that the constraints can be autoregressively satisfied.  

We have given explicit constructions of AR-NN which exactly represent the ground and excited states of several models, including the 2D and 3D toric codes as well as the X-cube fracton model. For those models for which exact representations are unknown, 
we variationally optimize our symmetry incorporated AR-NN to obtain either high-quality ground states or time-dependent wave functions.  This has been done for the U(1) quantum link model, $\mathbb{Z}_N$ gauge theory, the $\text{SU(2)}_3$ anyonic chain, and the SU(2) quantum spin-$1/2$ chain.  For these systems we are able to measure dynamical properties, produce phase diagrams, and compute observables accurately. 

Our approach opens up the possibility of probing a larger variety of models and the physics associated with them.  For example, the higher spin representation $S > 1/2$ in the (1+1)D QLM models would allow one to probe the quantum chromodynamics related physics of confinement and string breaking~\cite{PhysRevLett.109.175302}. For the (3+1)D QLM models, there is the Coulomb phase which manifests in pyrochlore spin liquids~\cite{Wiese_2013}. For $\mathbb{Z}_N$ gauge theory, it will be interesting to consider the general $\mathbb{Z}_N$ toric code with transverse field or disorder, with a goal of understanding its phase diagram. Recently, there have been proposals to understand the $\mathbb{Z}_N$ X-cube fracton model with non-trivial statistical phases~\cite{Shirley_2019}. For non-abelian anyons, the general $\text{SU(2)}_k$ formulation exhibits rich physics for different $k$ and one can study the corresponding topological liquids and edge states~\cite{Gils_2009}. Our approach can also be extended to study the phase diagram for the 2D Heisenberg models with SU(2) symmetry. 

Besides exploring various models in condensed matter physics and high energy physics, our approach can also be further applied to quantum information and quantum computation. Fibonacci anyons are known to support universal topological quantum computation, which is robust to local perturbations~\cite{Nayak_2008}. It will be interesting to see how well one can approximately simulate topological quantum computation or different braiding operations with anyonic symmetric autoregressive neural networks. As toric codes are an important example of quantum error correction code, our approach can be used to approximately study the performance of a toric code under different noise conditions. With respect to the recent efforts on simulating lattice gauge theories with quantum computation, our approach also provides an alternative method to compare to and benchmark quantum computers. In summary, the approach we have developed is versatile and powerful for investigating condensed matter physics, high energy physics and quantum information science. 

\section*{Acknowledgement}
D.L. is grateful for insightful discussion in high-energy physics with J. Stokes and J. Shen. D.L. acknowledges helpful
discussion with L. Yeo, O. Dubinkin, R. Levy, P. Xiao, R. Sun, and G. Carleo. This work is supported by the National Science Foundation under Cooperative Agreement No. PHY2019786 (the NSF AI Institute for Artificial Intelligence and Fundamental Interactions http://iaifi.org/). This work utilizes resources supported by the National Science Foundation’s Major Research Instrumentation program, Grant No. 1725729, as well as the University of Illinois at Urbana-Champaign~\cite{hal}. The authors acknowledges MIT Satori and MIT SuperCloud \cite{reuther2018interactive} for providing HPC resources that have contributed to the research results reported within this paper. Z.Z. is partially supported by NSF DMS-1854791, NSF OAC-1934757, and Alfred P. Sloan Foundation. Vera Mikyoung Hur is partially supported by NSF DMS-1452597 and DMS-2009981. This work is supported in part by the U.S. Department of Energy, Office of Science, Office of High Energy Physics QuantISED program under an award for the Fermilab Theory Consortium "Intersections of QIS and Theoretical Particle Physics”.

\newpage

\bibliography{reference}

\newpage
\clearpage

\begin{appendix}

\section{Additional Results for Quantum Link Model and Toric Code Model}\label{sec:additional results}

\begin{figure}[ht!]
    \centering
    \includegraphics[width=0.95\linewidth]{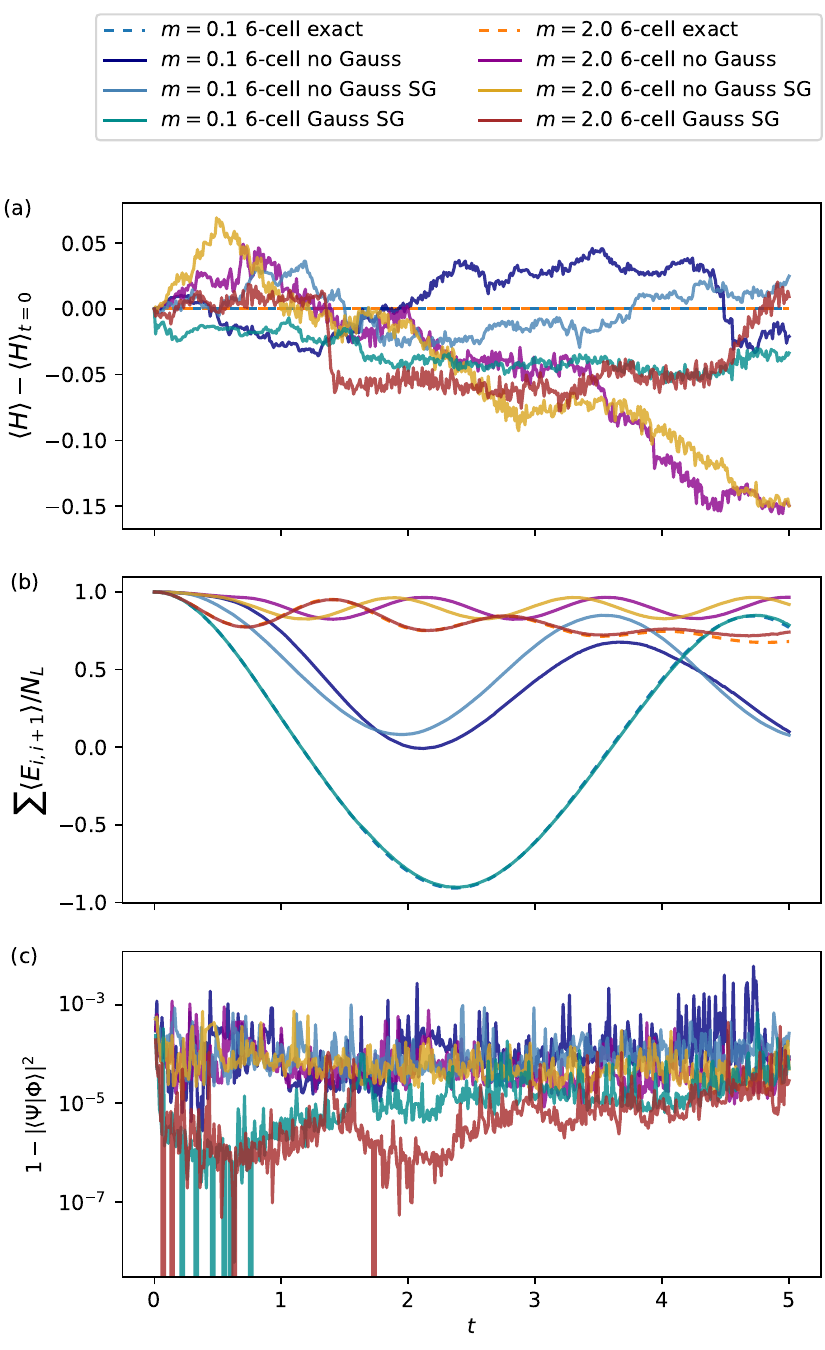}
    \caption{Dynamics for the 6-unit-cell (12 sites and 12 links) open-boundary QLM for $m=0.1$ and $m=2.0$ with and without gauge invariant construction. The dashed curves are the exact results from the exact diagonalization for 6 unit cells. The ``6-cell Gauss SG'' is the same as the ``6-cell Gauss'' in Fig.~\ref{fig:qlm_dynamics}. (a) The change in the energy during the dynamics. (b) The expectation value of the electric field averaged over all links. (c) The per step infidelity measure, where $\ket{\Psi}$ and $\ket{\Phi}$ are defined in Sec.~\ref{sec:optimization}.
    We use the Transformer neural network with 1 layer, 16 hidden dimensions and the real-imaginary parameterization (see Fig.~\ref{fig:parameterization}). The initial state is $\ket{\bullet\rightarrow\circ\rightarrow}$ for each unit cell and we train the neural network using the forward-backward trapezoid method with the time step $\tau = 0.005$, 600 iterations in each time step, and 12000 samples in each iteration. For the results labeled with SG, we used the sign gradient (SG) optimizer \cite{Otis_2019} for 15-30 iterations (depending on the resulting fidelity) before switching to the regular optimizer. The neural network architecture, initialization and optimization details are discussed in Appendix~\ref{app:nn}.}  
    \label{fig:qlm_dynamics_6}
\end{figure}

\begin{figure}[ht!]
    \centering
    \includegraphics[width=\linewidth]{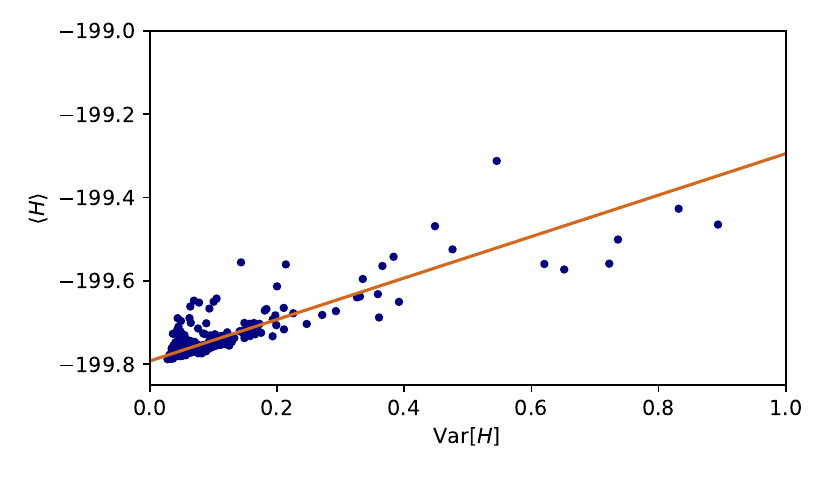}
    \caption{Ground state energy extrapolation at $m=0.7$ for 160 unit cells. The energies and variances are obtained from the training output of RNN in Fig.~\ref{fig:qlm_variational_fields}. We used a linear fit for variances smaller than 0.2 and obtained a y-intercept of $-199.7923$. Our variational ground state energy is $-199.7803\pm0.0005$. The ansatz, initialization and optimization are the same as in Fig.~\ref{fig:qlm_variational_fields} and are discussed in Appendix~\ref{app:nn}. }
    \label{fig:qlm_extrapolation}
\end{figure}

\begin{figure}[ht!]
    \centering
    \includegraphics[width=\linewidth]{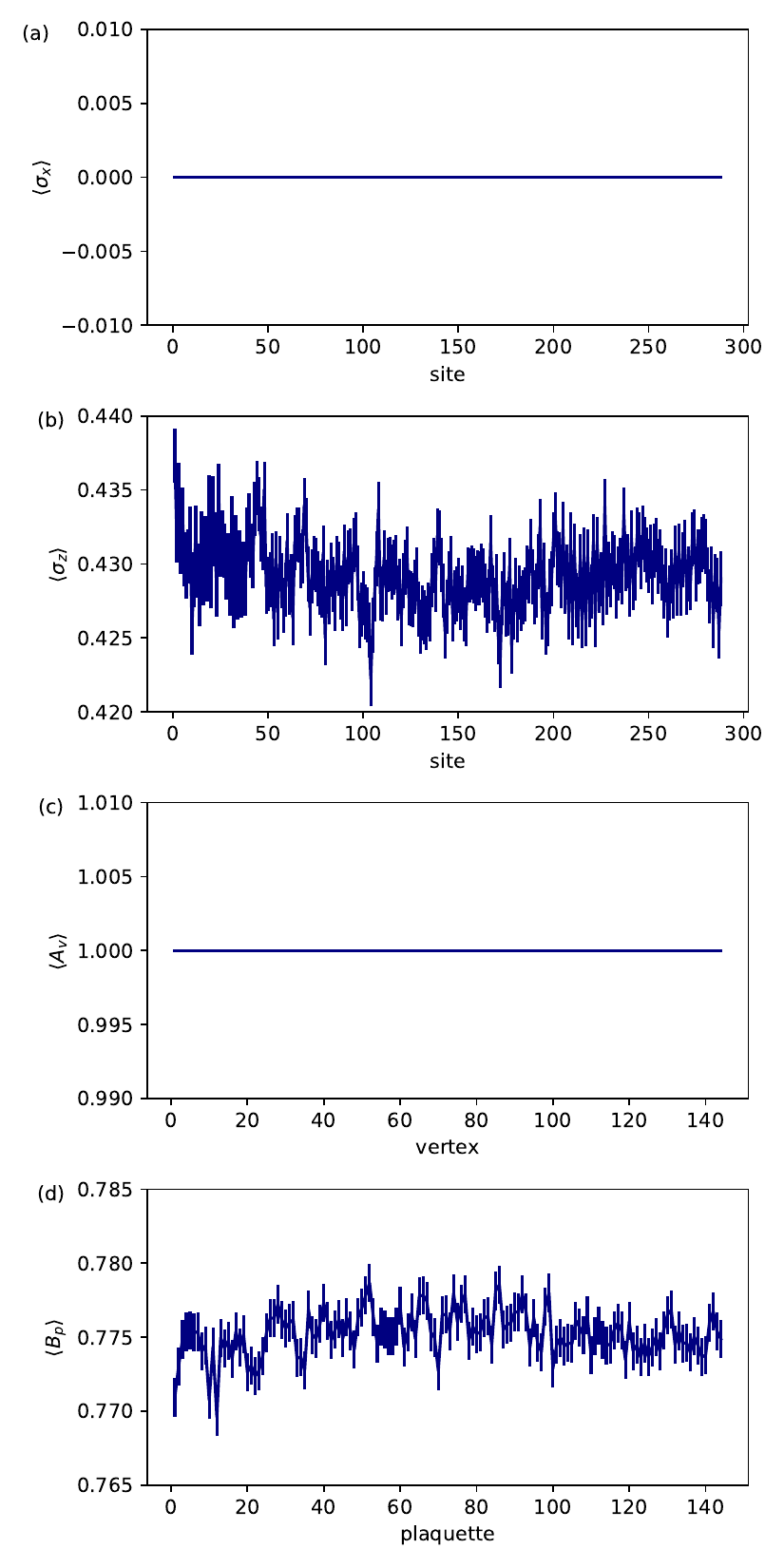}
    \caption{Local observables: (a) Expectation value of $\ev{\sigma_x}$, (b) $\ev{\sigma_z}$, (c) vertex operator $\ev{A_v}$  and (d) plaquette operator $\ev{B_p}$ (defined in Eq.~\ref{eq:2d_toric}) for the $12\times12$ toric code model with $h=0.36$. The neural network is the same as in Fig.~\ref{fig:toric_code_compare}. The neural network architecture, initialization and optimization details are discussed in Appendix~\ref{app:nn}.} 
    \label{fig:toric_code_local_obs}
\end{figure}

\begin{figure}[ht!]
    \centering
    \includegraphics[width=\linewidth]{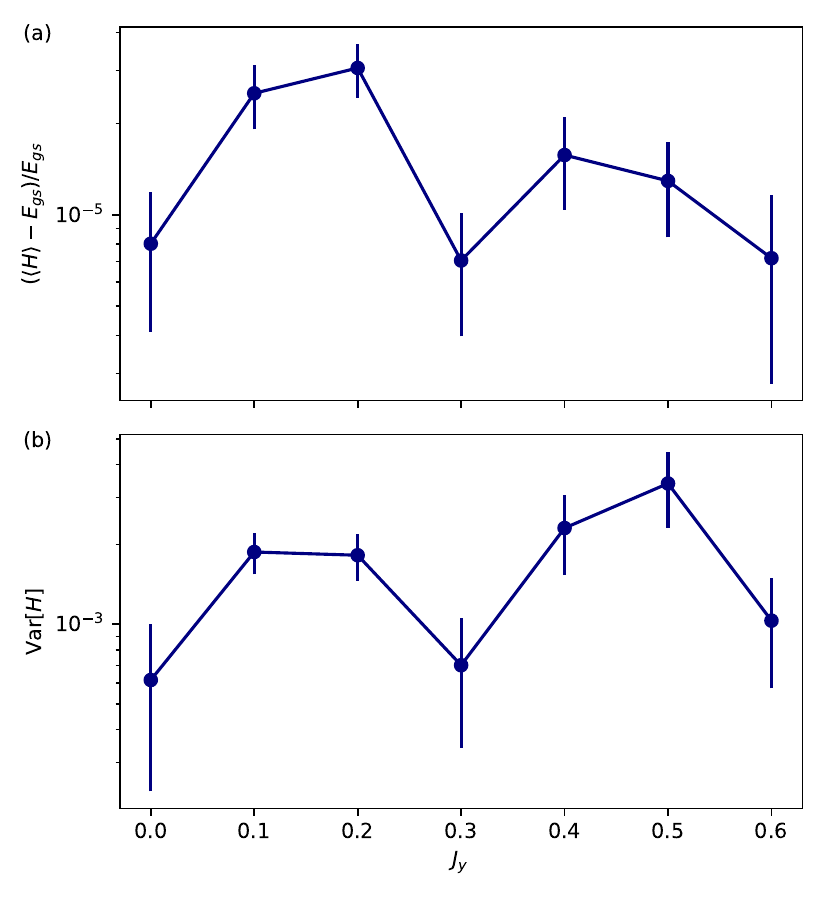}
    \caption{ (a) Relative error in energy and (b) variance of energy for $3\times3$ toric code model with an additional term discribed in Eq.~\ref{eq:toric_transverse}. Here we choose $h=0.36$ and run the neural network for different $J_y$'s. We use the 2D RNN neural network with real-imaginary parameterization. The neural network is trained using Adam optimizer for 10000 iteraions with 12000 samples in each iteration. The neural network architecture, initialization and optimization details are discussed in Appendix~\ref{app:nn}.} 
    \label{fig:3x3_toriccode_y}
\end{figure}

\begin{figure}[ht!]
    \centering
    \includegraphics[width=\linewidth]{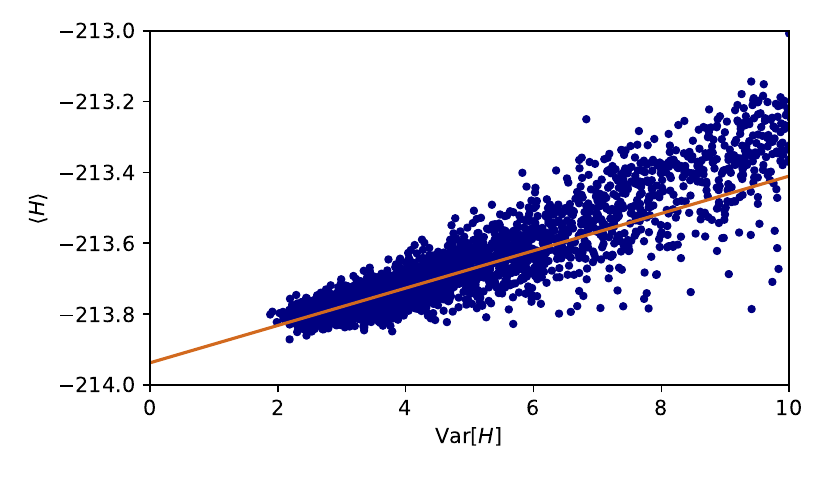}
    \caption{ Ground state energy extrapolation at $h=0.36$ and $J_y=0.3$ for $10\times10$ modified toric code model. The energies and variances are obtained from the training output. We used a linear fit for variances smaller than 0.2 and obtained a y-intercept of $-213.938$. Our variational ground state energy is $-213.802\pm0.003$. We use the 2D RNN neural network with real-imaginary parameterization. The neural network is trainined using the transfer learning technical for 5000 iterations after the $3\times3$ result with 12000 samples in each iteration. The neural network architecture, initialization and optimization details are discussed in Appendix~\ref{app:nn}.} 
    \label{fig:extrapolation_10x10}
\end{figure}

In this section, we present additional results. Fig.~\ref{fig:qlm_dynamics_6} is a 6-unit-cell quantum link model dynamics, which we can exactly diagonalize. We observed that the gauge invariant AR-NN matches the exact results up to $t=4$, while the non-gauge ansatz quickly fails to capture the electric fields even with the sign gradient (SG) optimizer \cite{Otis_2019} (Fig.~\ref{fig:qlm_dynamics_6} (b)). In addition, the gauge invariant AR-NN in general has a lower per step infidelity (Fig.~\ref{fig:qlm_dynamics_6}(c)).

In Fig.~\ref{fig:toric_code_local_obs}, we measure the local observables of the $12\times 12$ toric code model. We show that even though the neural network does not automatically preserve translational symmetry, the optimization drives the neural network to a translationally symmetric state. In addition, the weights of RNN is translationally invariant, which, although not guaranteed, could be potentially useful for preserving translational symmetry.

 Furthermore, we benchmark our method on the following model:

\begin{equation}\label{eq:toric_transverse}
    H = -\sum_{v\in V} A_v -\sum_{p\in P} B_p - h \sum_{e \in E} \sigma^z_e - j_y \sum_{p\in P} \prod_{e \in p} \sigma^y_e,
\end{equation}

With the additional $\sum_{p\in P} \prod_{e \in p} \sigma^y_e$, the Hamiltonian exhibits a sign problem compared to the original toric model, which would be challenging for the Monte Carlo method. We further test our method first on small systems ($3 \times 3$ to compare against exact diagonalization) and find relative energy differences on the order of $10^{-5}$ suggesting good agreement (see Fig.~\ref{fig:3x3_toriccode_y}).  We further apply our method on a $10 \times 10$ lattice.  Though it's not clear what to benchmark exactly against here, we measure the energy difference from a variance extrapolated result, and find our variational answer is close to it within $\Delta E \sim 0.1$ (see Fig.~\ref{fig:extrapolation_10x10}).

\section{Kitaev's $D(G)$ Model and \\Exact Representation of Ground State}\label{app: DG_model}

\begin{figure}[h]
        \includegraphics[scale=0.5]{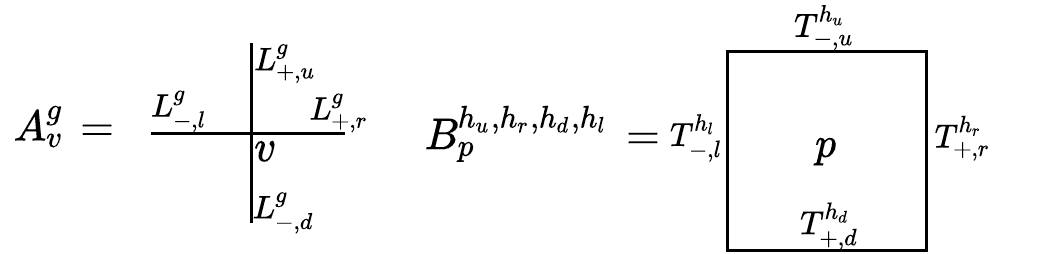}
   \caption{\label{fig:operator} $A_v^g$ and $B_p^{h_u, h_r, h_d, h_l}$ operators. $A_v^g = L_{+, u}^g L_{+, r}^g L_{-, d}^g L_{-, l}^g$ and $B_p^{h_u, h_r, h_d, h_l} = T_{-, u}^{h_u} T_{+, r}^{h_r} T_{+, d}^{h_d} T_{-, l}^{h_l} $, where $L_{+}^g\ket{z} = \ket{gz}$, $L_{-}^g\ket{z}=\ket{zg^{-1}}$, $T_{+}^h \ket{z} = h\delta_{h, z}\ket{z}$ and $T_{-}^h \ket{z} = h^{-1}\delta_{h, z} \ket{z}$.}
\end{figure}

We generalize our gauge invariant autoregressive construction for the 2D $\mathbb{Z}_2$ toric code.
Kitaev's $D(G)$ model~\cite{Kitaev_2003} is defined on an $L \times L$ periodic square lattice where each edge has a basis $\{\ket{g}, g \in G\}$ for some group $G$. Here we focus on finite groups, in particular $G = \mathbb{Z}_N$ for $\mathbb{Z}_N$ theory. Without loss of generality, we attach an upward arrow for each edge in the $y$-direction and a right arrow for each edge in the $x$-direction. We employ the notation of Sec.~\ref{sec:2D toric} and introduce operators $A_v^g$ and $B_p^{h_u, h_d, h_l, h_r}$ as in Fig.~\ref{fig:operator}.

The Hamiltonian defined on $\mathcal{H} (G)^{\otimes E}$ is 
\begin{equation}
    H = -\sum_{v \in V} A_v- \sum_{p \in P} B_p,
\end{equation}
where $A_v = \frac{1}{|G|} \sum_{g \in G} A_v^g$ is Gauss's law and the gauge constraint, and $B_p = \sum_{h_u h_r h_d h_l = \mathbbm{1}_G} B_p^{h_u, h_r, h_d, h_l}$. 

Let $\ket{+} = \frac{1}{\sqrt{|G|}} \sum_{g \in G} \ket{g}$, and $\psir = \prod_{p \in P} B_p \ket{+}^{\otimes E}$ is the ground state. This is because $\psir$ is a ground state for each $A_v$ and $B_p$. It is easy to verify that $B_p \psir = \psir$. To see $A_v \psir =\psir$, notice that $A_v$ and $B_p$ commute 
and $A_v \ket{+}^{\otimes E} =\ket{+}^{\otimes E}$. Similarly to the $\mathbb{Z}_2$ toric code, the ground state can be constructed using gauge invariant autoregressive neural networks by defining each star as a composite particle and checking Gauss's law and bond consistency.

\section{(1+1)D $\mathbb{Z}_N$ Lattice QED Model}\label{app:zd_qed}

Our approach in Sec.~\ref{sec:QLM} can be applied to the (1+1)D $\mathbb{Z}_N$ lattice quantum electrodynamics (QED) model, which is a discretization  of the Schwinger model for the continuous-space QED in 1+1 dimensions~\cite{Notarnicola_2015}. The (1+1)D $\mathbb{Z}_N$ model takes a similar form as the (1+1)D QLM, which has fermions on sites and electric fields on links between two sites. Let $\{ |e_{i,i+1}\rangle \}$ for $1 \leq e_{i,i+1} \leq N$ denote the orthonormal basis on each link $(i,i+1)$. The (1+1)D $\mathbb{Z}_N$ gauge theory can take the following form~\cite{Notarnicola_2015}
\begin{equation}
\begin{aligned}
H = & - \sum _ {i  } \left[ \psi _ { i } ^ { \dagger } U _ { i , i + 1 } \psi _ { i + 1 } + \psi _ { i + 1 } ^ { \dagger } U _ { i , i + 1 } ^ { \dagger } \psi _ { i } \right] \\
& + m \sum _ { i } ( - 1 ) ^ { i } \psi _ { i } ^ { \dagger } \psi _ { i } 
+ \frac { g ^ { 2 } } {8} \sum _ { i } (V _ { i , i + 1 } - \mathbbm{1})(V^{\dagger}_{ i , i + 1 } - \mathbbm{1}),
\end{aligned}
\label{eq:Zd_qed}
\end{equation}
where $U_{i,i+1}\ket{e_{i,i+1}} = \ket{ (e_{i,i+1}+1) \text{mod } N}$, and $V_{i,i+1}|e_{i,i+1}\rangle = e^{-i2\pi m/N}|e_{i,i+1}\rangle$ for $m=e_{i,i+1}$. The Gauss's law operator $G_i$ of the model can be written as
\begin{equation}
    G_i = e^{\frac{i2\pi}{N} (\psi^{\dagger}_i \psi_i + \frac{1}{2}(-1)^{i} - \frac{1}{2}) } V_{i,i+1} V_{i-1,i}^{\dagger}
\label{eq:zd_gauss}
\end{equation}
such that $G_i \psir = \psir$ for each $i$~\cite{Notarnicola_2015}.

Similarly to the (1+1)D QLM, one can construct the gauge invariant autoregressive neural network as Fig.~\ref{fig:qlm} and perform gauge checking with $G_i$ in Eq.~\ref{eq:zd_gauss}.

\section{$\text{SU}(2)$ Invariant Autoregressive \\Neural Network for Heisenberg Model}\label{app:su2}

The 1D Heisenberg Model is described as
\begin{equation}
    H = \sum_i \sigma^x_i \sigma^x_{i+1} + \sigma^y_i \sigma^y_{i+1} + \sigma^z_i \sigma^z_{i+1}.
\end{equation}
We work in the angular momentum basis $\{\ket{j_1, j_2, j_3, \dots, j_n}\}$ similarly to ~\cite{Gils_2009}, instead of the spin basis, to construct an $\text{SU}(2)$ invariant autoregressive wave function. Each $j_i$ is the total angular momentum quantum number for spins from $1$ to $i$ and $j_n\equiv J$ is the total angular momentum quantum number for all spins. For the ground state of the Heisenberg model, the total angular momentum is zero, so $j_n=0$. We define the first composite particle as $j_1$ and the $i$'th composite particle as the
difference $j_i-j_{i-1}$. Note that this uniquely defines a physical state.  We then autoregressively enforce $j_{i<n}\ge0$ and $j_n=0$ as gauge checking, to achieve the $\text{SU}(2)$ invariant property.

\begin{figure}[h]
    \centering
    \includegraphics[width=\linewidth]{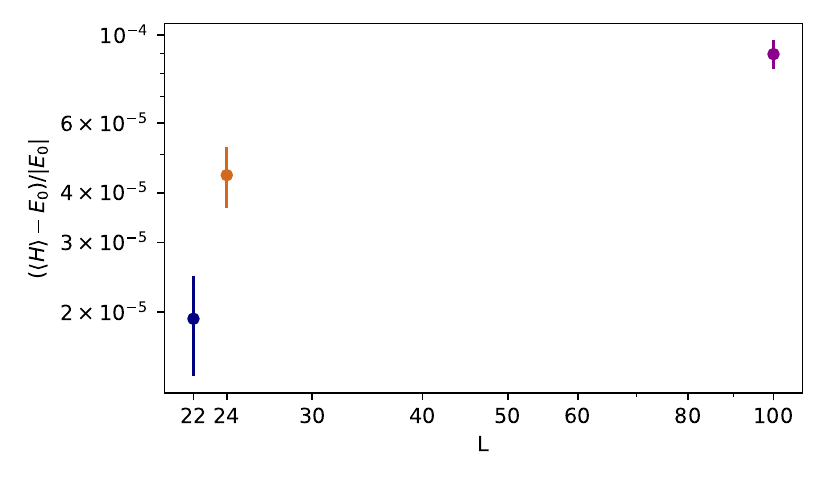}
    \caption{Relative error of variational ground state energy for the 1D Heisenberg model with $\text{SU}(2)$ symmetry for 22, 24 and 100 spins. We use the 1D RNN with 3 layers, $L$ hidden dimensions and the real-imaginary parameterization. We train the neural network for 5000 iterations with 12000 samples in each iteration. The exact solutions for 22 spins and 24 spins are computed with the exact diagonalization, and the exact solution for 100 spins is the DMRG result in Ref.~\onlinecite{rnn_wavefunction}. }  
    \label{fig:su2_heisenberg}
\end{figure}

Fig.~\ref{fig:su2_heisenberg} demonstrates the performance of our $\text{SU}(2)$ invariant autoregressive neural network on the Heisenberg model with $\text{SU}(2)$ symmetry.

\section{Neural Network Architecture} \label{app:nn}

\subsection{Complex Parameterization}

\begin{figure}[h]
    \centering
    \includegraphics[scale=0.5]{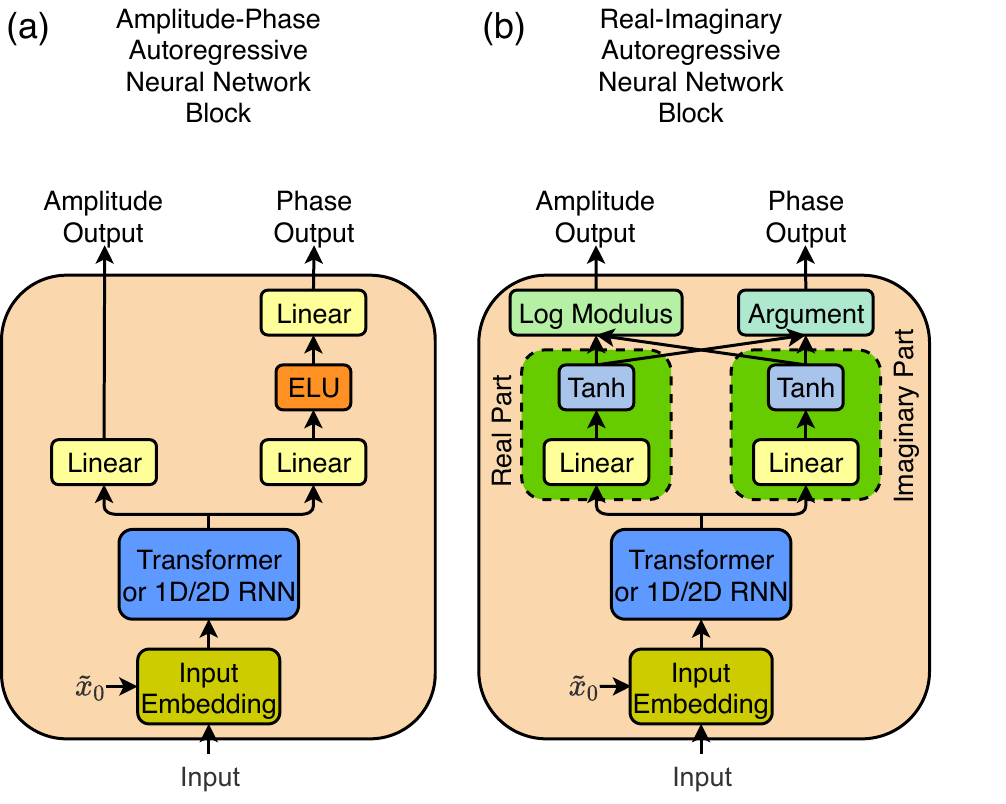}
    \caption{Two parameterizations of complex wave functions from autoregressive neural networks. (a) The amplitude-phase parameterization. The raw output is used as the input of both the amplitude branch and the phase branch. (b) The real-imag parameterization. The raw output is used as the input of both the real branch and the imaginary branch, which later is converted to the amplitude branch and the phase branch. }  
    \label{fig:parameterization}
\end{figure}

Wave functions are complex in general but both the Transformer network and 1D/2D RNN are real. We use two approaches (Fig.~\ref{fig:parameterization})---(a) amplitude-phase and (b) real-imaginary---to parametrize complex wave functions from real neural networks. 
In both parameterizations, the input configuration $\bm{\widetilde{x}}$, together with a default configuration $\widetilde{x}_0$, is embedded (i.e. each state of a composite particle is mapped to a unique vector) before fed into the Transformer or 1D/2D RNN. Certain gauge blocks in an AR-NN  take a default state $\widetilde{x}_0$ as opposed to any state of the composite particles; the embedded vector of this default state has arbitrary parameters which are trained during optimization. 

\subsection{Transformer}

\begin{figure}[H]
    \centering
    \includegraphics[scale=0.5]{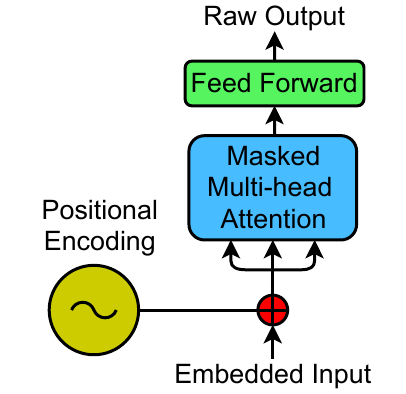}
    \caption{A single layer Transformer network. The embedded input is fed into the Transformer and the positional encoding is added. After a masked multi-head self-attention is applied, a feed forward layer produces the raw output.}  
    \label{fig:transformer}
\end{figure}

The Transformer used in this work (Fig.~\ref{fig:transformer}) is the same as the Transformer used in Ref.~\onlinecite{quantum_circuit}, which can be viewed as the standard Transformer encoder with masked multi-head attention from Ref.~\onlinecite{transformer} but without an additional add \& norm layer. The Transformer consists of a standard positional encoding layer, which uses sinusoidal functions to encode the positional information of the embedded input. After positional encoding, the input is fed into the standard masked multi-head attention mechanism. The mask here is crucial for autoregressiveness, as it only allows each site to depend on the previous sites. The output of the attention layer is then passed through a standard feed forward layer. The detailed explanation of the Transformer can be found in Refs.~\onlinecite{transformer, quantum_circuit}. This transformer is essentially equivalent to the standard PyTorch implementation~\cite{paszke2019pytorch}, but was implemented independently because that implementation did not exist at the start of our work.

\subsection{RNN Cells}

\begin{figure}[h]
    \centering
    \includegraphics[scale=0.5]{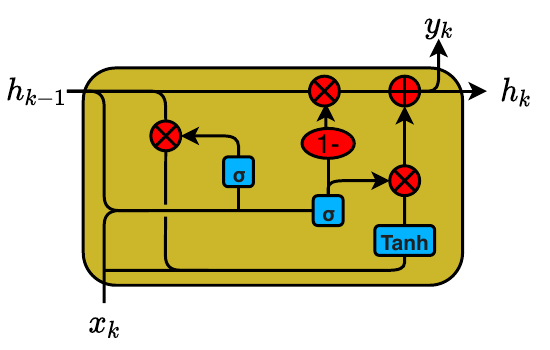}
    \caption{The GRU cell~\cite{gru} on which different RNNs are constructed. This is the same GRU cell as the PyTorch~\cite{paszke2019pytorch} implementation. }  
    \label{fig:grucell}
\end{figure}

For all RNNs in this work, we used the gated recurrent unit (GRU) cell~\cite{gru} (Fig.~\ref{fig:grucell}) in PyTorch~\cite{paszke2019pytorch}, which takes one input vector $x_k$, the hidden input $h_{k-1}$, and computes
\begin{equation}
\begin{aligned}
    &r = \sigma(W_{xr}x_{k} + b_{xr} + W_{hr} h_{k-1} + b_{hr} ),\\
    &z = \sigma(W_{xz}x_{k} + b_{xz} + W_{hz} h_{k-1} + b_{hz} ),\\
    &n = \tanh(W_{xn}x_{k} + b_{xn} + r \odot (W_{hn}h_{k-1} + b_{hn})),\\
    &h_k = (1 - z) \odot n + z \odot h_{k-1}, \\
    &y_k = h_k,
\end{aligned}
\end{equation}
where $\sigma$ is the sigmoid function and $\odot$ means element-wise product.

\begin{figure}[h]
    \centering
    \includegraphics[scale=0.5]{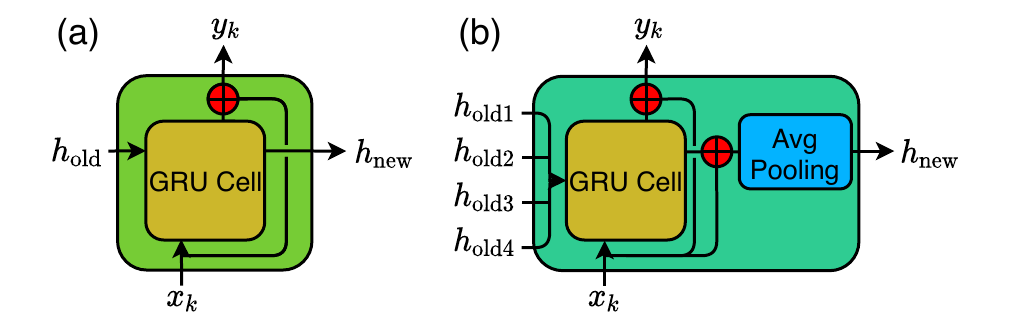}
    \caption{(a) 1D RNN cell. A ResNet (skip connection)~\cite{resnet} is added between the input $x_k$ and output $y_k$. To be noted that $x_k$, $y_k$, $h_\text{old}$ and $h_\text{new}$ has the same dimension. (b) 2D RNN cell with periodic boundary condition. This cell requires four hidden inputs $h_{\text{old}i,i=1,2,3,4}$ and generates one hidden output $h_\text{new}$. Skip connections are added for both output $y_k$ and hidden output $h_\text{new}$. The average pooling reduces the hidden output to have the same dimension as each hidden input. To be noted that the dimension of $x_k$ and $y_k$ is four times the dimension of each $h_{\text{old}i}$ and $h_\text{new}$.}  
    \label{fig:rnncell}
\end{figure}

We then build 1D and periodic 2D RNN cells (Fig.~\ref{fig:rnncell}) based on the GRU cell. The 1D RNN cell computes
\begin{equation}
\begin{aligned}
    & (y_\text{raw}, h_\text{new}) = \text{GRUcell}(x_k, h_\text{old}),\\
    & y_k = y_\text{raw} + x_k,
\end{aligned}
\end{equation}
whereas the periodic 2D RNN cell computes
\begin{equation}
\begin{aligned}
    & (y_\text{raw}, h_\text{raw}) = \text{GRUcell}(x_k, [h_\text{old1}, h_\text{old2}, h_\text{old3}, h_\text{old4}]),\\
    & [h_\text{new1}, h_\text{new2}, h_\text{new3}, h_\text{new4}] = h_\text{raw} + x_k, \\
    & h_\text{new} = \frac{1}{4}\left(h_\text{new1}+ h_\text{new2}+h_\text{new3}+ h_\text{new4}\right), \\
    & y_k = y_\text{raw} + x_k.
\end{aligned}
\end{equation}

\subsection{RNNs}
With the RNN cells, we can build 1D and periodic 2D RNNs.

\begin{figure}[h]
    \centering
    \includegraphics[scale=0.5]{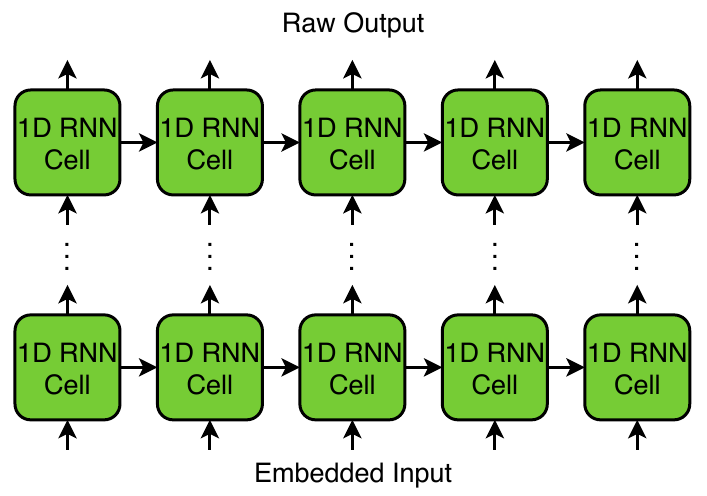}
    \caption{1D RNN built from 1D RNN Cells. The neural network has a multi-layer design similar to Pytorch GRU implementation~\cite{paszke2019pytorch,gru}. The weight matrices and biases are shared between different layers.}  
    \label{fig:1drnn}
\end{figure}

The 1D RNN (Fig.~\ref{fig:1drnn}) has a multi-layer design and shares the same structure as the Pytorch~\cite{paszke2019pytorch} GRU~\cite{gru}. The embedded input configuration is fed into the cells one at a time through multiple layers and produces a raw output. In our work, the cells at different layers share the weight matrices and bias vectors.

\begin{figure}[h]
    \centering
    \includegraphics[scale=0.5]{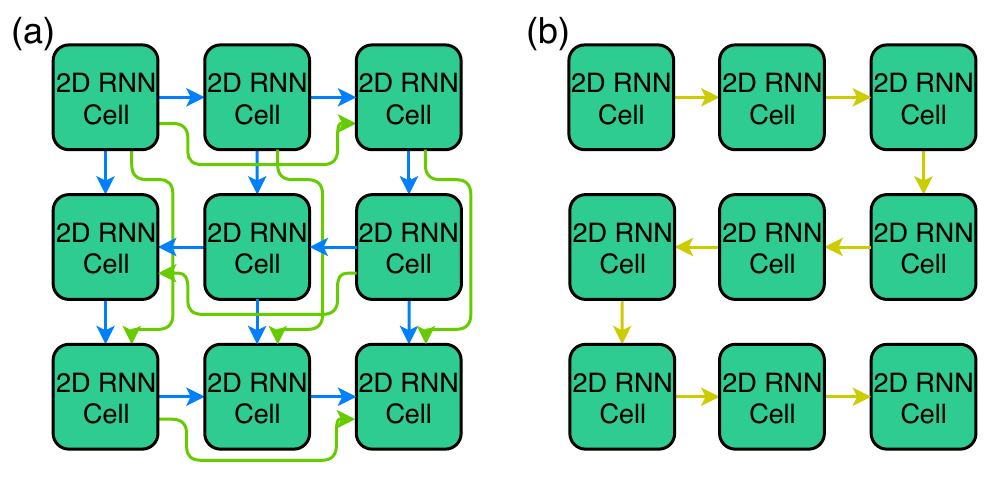}
    \caption{(a) The hidden information path for 2D RNN with periodic boundary condition for a $3\times3$ system. Blue arrows show the non-boundary information path  whereas the green arrows show the periodic boundary information path. When there are less than four hidden inputs, zero vectors are used for padding. (b) The conditioning and sampling order of the 2D RNN.}  
    \label{fig:2drnnlayer}
\end{figure}

The periodic 2D RNN has a more complicated design to capture the most correlations and can be viewed as a periodic extension of the 2D RNN in Ref.~\onlinecite{rnn_wavefunction}. In each layer, the hidden vector $h$ is passed around according to Fig.~\ref{fig:2drnnlayer}(a), where each cell receives a maximum number of four hidden vectors and concatenates them according to Fig.~\ref{fig:rnncell}(b). When the number of hidden vectors received is less than four, zero vectors are used to pad the concatenated hidden vector to the correct length. The configuration is evaluated and sampled in a zigzag $S$ path (Fig.~\ref{fig:2drnnlayer}(b)) to ensure autoregressiveness.

\begin{figure}[h]
    \centering
    \includegraphics[scale=0.5]{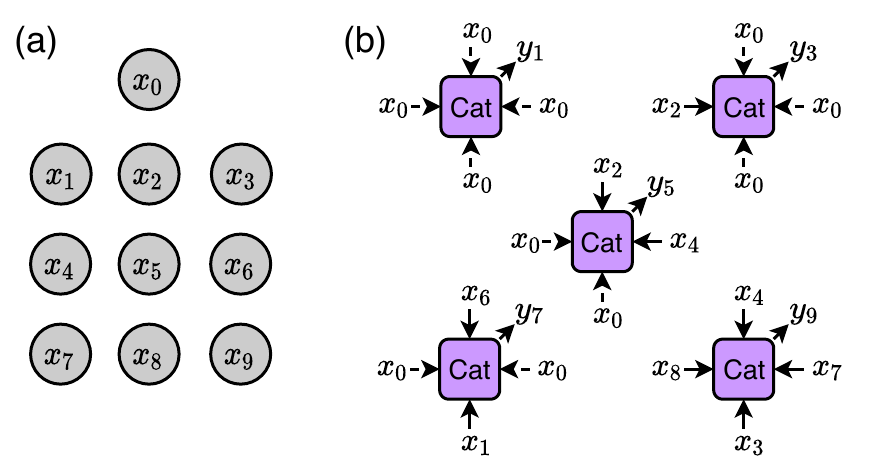}
    \caption{2D RNN input concatenation layer. For (a) a $3\times3$ input array with a default input $x_0$, (b) the concatenation layer takes the four input vectors surrounding each site with periodic boundary condition and outputs the concatenated vector of the four surrounding vectors. $x_0$ is used when the surrounding inputs appear later in the conditioning order.}  
    \label{fig:2drnncat}
\end{figure}

Before the first layer of the periodic 2D RNN, a special concatenation of the embedded input needs to be performed. At each location, the concatenation layer takes the four surrounding inputs (periodically) and concatenates them into a single vector. If any (or all) of the surrounding inputs lies later in the conditioning order in Fig.~\ref{fig:2drnnlayer}(b), the corresponding input is replaced with a default input $x_0$. For a $4\times4$ 2D input array shown in Fig.~\ref{fig:2drnncat}(a), some concatenation examples are shown in Fig.~\ref{fig:2drnncat}(b). After the first layer, the output of a previous layer can be directly fed into the next layer similar to a regular RNN without any further process.

\begin{figure}[h]
    \centering
    \includegraphics[scale=0.5]{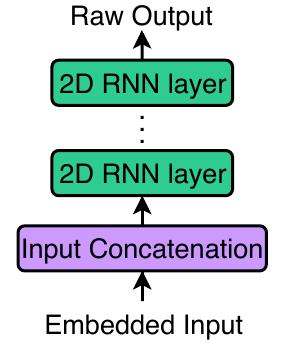}
    \caption{The 2D RNN is built from one input concatenation layer and multiple 2D RNN layers. The weight matrices and biases are shared between different layers.}  
    \label{fig:2drnn}
\end{figure}

A multi-layer periodic 2D RNN consists of one input concatenation layer and several 2D RNN layers as shown in Fig.~\ref{fig:2drnn}.

\subsection{Initialization and Optimization}
We use different initialization techniques for different models. For the QLM model, the initialization is done through tomography, minimizing $-\log \abs{\psi(x)}^2$ for a desired configuration $x$ ($\ket{\bullet\rightarrow\circ\rightarrow}$ for each unit cell in this case) .For the 2D toric code model, we set the weight matrix in the last linear layer to be 0 and the bias such that the wave function for each composite particle is $\sqrt{0.23}\scriptstyle\ket{\scriptstyle0 \underset{\scriptstyle0}{\overset{\scriptstyle0}{{\color{white}\cdot}}} 0} + \displaystyle\sqrt{0.11}\scriptstyle\ket{\scriptstyle1 \underset{\scriptstyle0}{\overset{\scriptstyle1}{{\color{white}\cdot}}} 0} + \displaystyle\sqrt{0.11}\scriptstyle\ket{\scriptstyle1 \underset{\scriptstyle1}{\overset{\scriptstyle0}{{\color{white}\cdot}}} 0} + \displaystyle\sqrt{0.11}\scriptstyle\ket{\scriptstyle0 \underset{\scriptstyle1}{\overset{\scriptstyle0}{{\color{white}\cdot}}} 1} + \displaystyle\sqrt{0.11}\scriptstyle\ket{\scriptstyle0 \underset{\scriptstyle0}{\overset{\scriptstyle1}{{\color{white}\cdot}}} 1} + \displaystyle\sqrt{0.11}\scriptstyle\ket{\scriptstyle1 \underset{\scriptstyle0}{\overset{\scriptstyle0}{{\color{white}\cdot}}} 1} + \displaystyle\sqrt{0.11}\scriptstyle\ket{\scriptstyle0 \underset{\scriptstyle1}{\overset{\scriptstyle1}{{\color{white}\cdot}}} 0} + \displaystyle\sqrt{0.11}\scriptstyle\ket{\scriptstyle1 \underset{\scriptstyle1}{\overset{\scriptstyle1}{{\color{white}\cdot}}} 1}$, which empirically produces a very low initial energy. We used a transfer learning technique, where we first train our neural network on a $6\times6$ model before training it on the $10\times10$ model. When transferred to the larger system, the weight and bias in the last linear layer is dropped and replaced with the initialization scheme described above.
\begin{figure}[h]
    \centering
    \includegraphics[width=\linewidth]{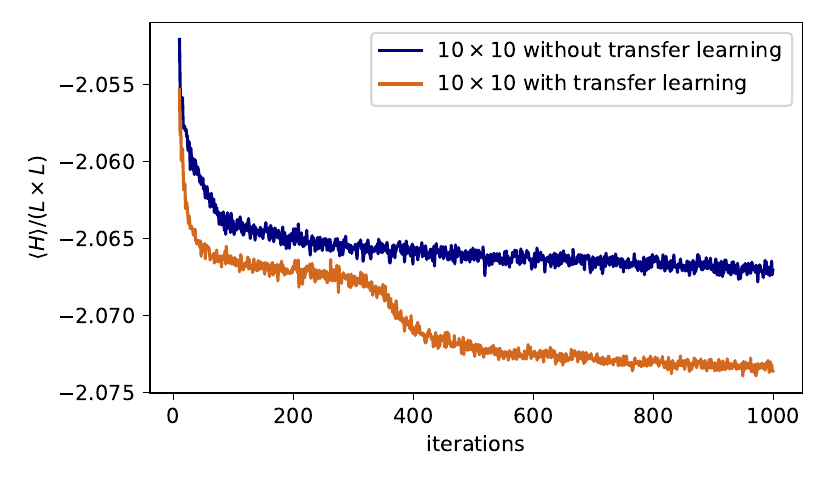}
    \caption{Per site energy with and without transfer learning during the training process for the  $10\times10$ toric code model. The first 10 iterations are not included as they are too large and outside the range of the figure. The energy with transfer learning is clearly lower than the energy without transfer learning.}  
    \label{fig:transfer}
\end{figure}
In Fig.~\ref{fig:transfer}, we show the effect of transfer learning on a $10\times10$ toric code model. With transfer learning, the energy is clearly lower than the energy without transfer learning.

For the anyon model, we set the weight matrix in the last linear layer to be 0 and the bias to be the state of $1/\sqrt{2}\ket{\mathbbm{1}} + 1/\sqrt{2}\ket{\tau}$ for each particle. When producing the phase diagram, we used a transfer learning technique, where we first train the neural network on 32 anyons with $\theta = 0, \pi/4, \pi/2, 3\pi/4, \pi, 5\pi/4, 3\pi/2, 7\pi/4, \text{and } 2\pi$ for 3000 iterations, and then transfer the model with $\theta$ that is closest to the desired value of $\theta$ for 40 anyons for another 3000 iterations. Similar to the toric code case, when transferred to the larger system, the weight and bias in the last linear layer is dropped and replaced by the initialization described above.  In all models, except the last layer, the weights and biases are initialized using PyTorch's~\cite{paszke2019pytorch} default initialization. 

For optimization, we used the Adam~\cite{adam} optimizer with an initial learning rate of 0.01. For the QLM dynamics, the learning rate is halved at iterations 100, 200, 270, 350, and 420, for the QLM ground state optimization; the learning rate is halved at iterations 300, 600, 900, 1200, 1800, 2400, 3000, 4000, 5000, 6000, and 7000; and for the ground state optimization of other models, the learning rate is halved at iterations 100, 500, 1000, 1800, 2500, 4000, and 6000. In addition, for the 6-unit-cell cases and 12-unit-cell $m=0.1$ case with Gauss's law, we use the sign gradient (SG) optimizer \cite{Otis_2019} for 15-30 iterations (depending on the resulting fidelity) before switching to the regular optimizer, and for the 12-unit-cell $m=2.0$ case with Gauss's law, we modified the loss function by adding an energy penalty term described in Appendix~\ref{app:ep}.

\subsection{Computational Complexity}

In this section, we explain the computational complexity of the neural networks used in this work.

In terms of scaling, the total cost per sweep is 
    \begin{itemize}
        \item  Transformer: $O(N^2 h^3)$ (evaluation complexity), $O(N^3 h^3)$ (sampling complexity);
        \item  RNN: $O(N h^2)$(computational complexity);
    \end{itemize}
where $h$ is the hidden dimension and $N$ is the size of the system. Note that the memory complexity
is bounded by the computational complexity. In the training process, we used 2-4 GPUs depending on the availability of GPUs in the cluster and never experienced any memory issue with a total of 64GB GPU memory. 
To give a sense for computational difficulty, generating Fig.~\ref{fig:toric_code_compare} takes six GPU days using Tesla V100 GPUs.  

\section{Energy Penalty} \label{app:ep}
\begin{figure}[h]
    \centering
    \includegraphics[width=0.95\linewidth]{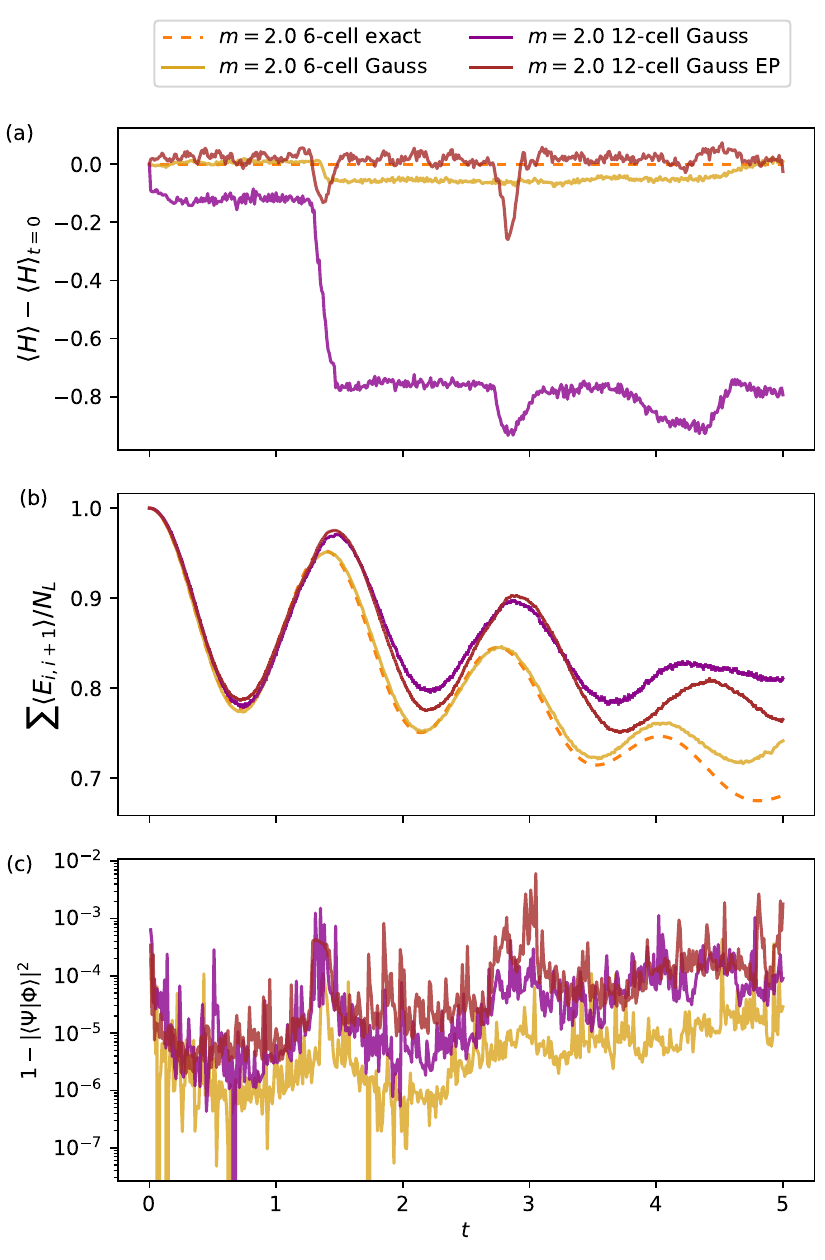}
    \caption{Dynamics for the 6- and 12-unit-cell (12-24 sites and 12-24 links) open-boundary QLM for $m=2.0$ with and without energy penalty. The dashed curves are the exact results from the exact diagonalization for 6 unit cells. The ``12-cell Gauss EP'' is the same as the ``12-cell Gauss'' in Fig.~\ref{fig:qlm_dynamics}. (a) The change in the energy during the dynamics. (b) The expectation value of the electric field averaged over all links. (c) The per step infidelity measure, where $\ket{\Psi}$ and $\ket{\Phi}$ are defined in Sec.~\ref{sec:optimization}.
    We use the Transformer neural network with 1 layer, 16 hidden dimensions for 6 unit cells and 32 hidden dimensions for 12 unit cells, and the real-imaginary parameterization (see Fig.~\ref{fig:parameterization}). The initial state is $\ket{\bullet\rightarrow\circ\rightarrow}$ for each unit cell and we train the neural network using the forward-backward trapezoid method with the time step $\tau = 0.005$, 600 iterations in each time step, and 12000 samples in each iteration. The neural network architecture, initialization and optimization details are discussed in Appendix~\ref{app:nn}.}  
    \label{fig:qlm_dynamics_ep}
\end{figure}
While quantum dynamics should exactly preserve the energy, as a practical matter when a variational state can't fully represent the exact dynamics, there can be a tension between maximizing fidelity per step and preserving the energy.  In some cases, it may be desirable to better preserve the total energy at some cost in fidelity per step.  Toward that end, we show one can introduce an additional term into the lost function which acts as a penalty toward the drift in energy. 
We demonstrate this for the $m=2.0$ 12-unit-cell QLM in Fig.~\ref{fig:qlm_dynamics}. The energy penalty term 
\begin{equation}
    \mathcal{L}_p =  \abs{\frac{1}{N}\sum_{x \sim \abs{\psi_\theta}^2}^N E_\text{loc}(x) - E_0}^2,
\end{equation}
where $E_\text{loc}(x)$ is defined in Eq.~\ref{eq:local_energy} and $E_0$ is the initial energy. 
This term is added to the dynamics loss function $\mathcal{L}_d$ (Eq.~\ref{eq:dynamics_loss}) to obtain the total loss function as
\begin{equation}
    \mathcal{L} = \mathcal{L}_d + \alpha \mathcal{L}_p,
\end{equation}
with $\alpha$ is a hyperparameter which we choose to be 0.01.
We show in Fig.~\ref{fig:qlm_dynamics_ep} this simulation with and without the energy penalty.  We find the dynamics as measured by the observables largely stays the same (and may be better), but the drift in the energy is significantly attenuated.

\section{Derivation of Stochastic Gradients \\for Variational and Dynamics Optimization} \label{app:gradient}
In Sec.~\ref{sec:optimization}, we presented stochastic gradients of the variational and dynamics optimizations. This section includes their derivations.

The variational optimization has been widely used and derived many times in other works~\cite{Carleo602, rnn_wavefunction}. Here we present the derivation for the sake of completeness:
\begin{align}
\begin{split}
    &\pdv{\ev{H}{\psi_\theta}}{\theta} \\ &= \mel{\pdv{\psi_\theta}{\theta}}{H}{\psi_\theta} + \mel{\psi_\theta}{H}{\pdv{\psi_\theta}{\theta}} \\
    &= 2 \sum_x \real \left\{\pdv{\psi_\theta^*(x)}{\theta} H \psi_\theta(x)\right\} \\
    &= 2 \sum_x \real \left\{\frac{1}{\psi_\theta^*(x)}\pdv{\psi_\theta^*(x)}{\theta}\psi_\theta^*(x) H \psi_\theta(x)\right\} \\
    &= 2 \sum_x \real \left\{\psi_\theta^*(x) H \psi_\theta(x) \pdv{}{\theta}\log\psi_\theta^*(x)\right\} \\
    &\approx \frac{2}{N} \sum_{x\sim\abs{\psi_\theta}^2}^N \real \left\{ \frac{H \psi_\theta(x)}{\psi_\theta(x)} \pdv{}{\theta}\log\psi_\theta^*(x)\right\} \\
    &\equiv \frac{2}{N} \sum_{x\sim\abs{\psi_\theta}^2}^N \real \left\{ E_\text{loc}(x) \pdv{}{\theta}\log\psi_\theta^*(x)\right\},
    \label{eq:before_var}
\end{split}
\end{align}
where the local energy is $E_\text{loc}(x) \equiv H \psi_\theta(x) / \psi_\theta(x)$. We can further control the variance by subtracting from the $E_\text{loc}(x)$ the average energy $E_\text{avg} \equiv \sum_{x\sim\abs{\psi_\theta}^2}^N E_\text{loc}(x) / N$ over the batch of samples~\cite{variance_reduction} as we did in Sec.~\ref{sec:optimization} and use the stochastic variance reduced gradient as
\begin{equation}
    \frac{2}{N} \sum_{x\sim\abs{\psi_\theta}^2}^N \real \left\{ \Big[E_\text{loc}(x) - E_\text{avg}\Big] \pdv{}{\theta}\log\psi_\theta^*(x)\right\},
\end{equation}
which has the same expectation as Eq.~\ref{eq:before_var}.

For the dynamics optimization gradient, as in Sec.~\ref{sec:optimization}, we define $\ket{\Psi_\theta} = (1 + iH\tau)\ket{\psi_{\theta(t + 2\tau)}}$ and $\ket{\Phi} = (1 - iH\tau)\ket{\psi_{\theta(t)}}$, and we drop $\theta(t)$ and name $\theta \equiv \theta(t + 2\tau)$. We start by splitting the negative log overlap:
\begin{equation}\label{eq:logsplit}
\begin{aligned}
    &-\log \frac{\ip{\Psi_\theta}{\Phi}\ip{\Phi}{\Psi_\theta}}{\ip{\Psi_\theta}\ip{\Phi}} \\&= -\log \ip{\Psi_\theta}{\Phi} - \log \ip{\Phi}{\Psi_\theta} + \log \ip{\Psi_\theta} + \log \ip{\Phi}.
\end{aligned}
\end{equation}
We then compute the gradient term by term. The first term on the right side of Eq.~\ref{eq:logsplit} becomes
\begin{align}
\begin{split}
    &\pdv{}{\theta} \log \ip{\Psi_\theta}{\Phi} \\&= \frac{1}{\ip{\Psi_\theta}{\Phi} } \ip{\pdv{\Psi_\theta}{\theta}}{\Phi} \\
    &= \frac{1}{\sum_x \Psi_\theta^*(x)\Phi(x)} \sum_x \pdv{\Psi_\theta^*(x)}{\theta}\Phi(x) \\
    &= \frac{1}{\sum_x \Psi_\theta^*(x)\Phi(x)} \sum_x \Psi_\theta^*(x)\Phi(x) \pdv{}{\theta}\log \Psi_\theta^*(x) \\
    &\approx \frac{1}{\sum_{x\sim{\abs{\psi_\theta}^2}}^N\frac{\Psi_\theta^*(x)\Phi(x)}{\abs{\psi_\theta(x)}^2}} \sum_{x\sim{\abs{\psi_\theta}^2}}^N \frac{\Psi_\theta^*(x)\Phi(x)}{\abs{\psi_\theta(x)}^2}\pdv{}{\theta}\log \Psi_\theta^*(x) \\
    &\equiv \frac{1}{N} \sum_{x\sim{\abs{\psi_\theta}^2}}^N \frac{\alpha(x)}{\alpha_\text{avg}}\pdv{}{\theta}\log \Psi_\theta^*(x),
\end{split}
\end{align}
where $\alpha(x) = \Psi_\theta^*(x)\Phi(x)/\abs{\psi_\theta(x)}^2$ and $\alpha_\text{avg} = \sum_{x\sim{\abs{\psi_\theta}^2}}^N \alpha(x)/N$ as in Sec.~\ref{sec:optimization}. The second term on the right side of Eq.~\ref{eq:logsplit} is just the complex conjugate of the first term, whereby
\begin{align}
\begin{split}
    \pdv{}{\theta} \log \ip{\Phi}{\Psi_\theta}
    =& \left[\pdv{}{\theta} \log \ip{\Psi_\theta}{\Phi}\right]^* \\
    \approx& \frac{1}{N} \sum_{x\sim{\abs{\psi_\theta}^2}}^N \left[\frac{\alpha(x)}{\alpha_\text{avg}}\pdv{}{\theta}\log \Psi_\theta^*(x)\right]^*.
\end{split}
\end{align}

The third term on the right side of Eq.~\ref{eq:logsplit} becomes
\begin{align}
\begin{split}
    &\pdv{}{\theta} \log \ip{\Psi_\theta} \\&= \frac{1}{\ip{\Psi_\theta} } \left(\ip{\pdv{\Psi_\theta}{\theta}}{\Psi_\theta} + \ip{\Psi_\theta}{\pdv{\Psi_\theta}{\theta}}\right)\\
    &= \frac{2}{\sum_x \abs{\Psi_\theta(x)}^2} \sum_x \real \left\{ \pdv{\Psi_\theta^*(x)}{\theta}\Psi_\theta(x)\right\} \\
    &= \frac{2}{\sum_x \abs{\Psi_\theta(x)}^2} \sum_x \real \left\{\abs{\Psi_\theta(x)}^2 \pdv{}{\theta}\log \Psi_\theta^*(x) \right\} \\
    &\approx \frac{2}{\sum_{x\sim{\abs{\psi_\theta}^2}}^N\frac{\abs{\Psi_\theta(x)}^2}{\abs{\psi_\theta(x)}^2}} \sum_{x\sim{\abs{\psi_\theta}^2}}^N \real \left\{ \frac{\abs{\Psi_\theta(x)}^2}{\abs{\psi_\theta(x)}^2}\pdv{}{\theta}\log \Psi_\theta^*(x) \right\} \\
    &\equiv \frac{2}{N} \sum_{x\sim{\abs{\psi_\theta}^2}}^N \real \left\{ \frac{\beta(x)}{\beta_\text{avg}}\pdv{}{\theta}\log \Psi_\theta^*(x) \right\},
\end{split}
\end{align}
where $\beta(x) = \abs{\Psi_\theta(x)}^2/\abs{\psi_\theta(x)}^2$ and $\beta_\text{avg} = \sum_{x\sim{\abs{\psi_\theta}^2}}^N \beta(x)/N$  as in Sec.~\ref{sec:optimization}. The last term on the right side of Eq.~\ref{eq:logsplit} is $\theta$ independent such that
\begin{equation}
    \pdv{}{\theta} \log \ip{\Phi} = 0.
\end{equation}

Combining all the terms together, 
\begin{align}
\begin{split}
    &\pdv{\theta}\left(-\log \frac{\braket{\Psi_\theta}{\Phi}\braket{\Phi}{\Psi_\theta}}{\braket{\Psi_\theta}{\Psi_\theta}\braket{\Phi}{\Phi}}\right) \\
    &\approx \frac{2}{N}\sum_{x \sim \abs{\psi_{\theta}}^2}^N \real\left\{\left[\frac{\beta(x)}{\beta_\text{avg}} - \frac{\alpha(x)}{\alpha_\text{avg}}\right]\pdv{\theta} \log\Psi_\theta^*(x)\right\}.
\end{split}
\end{align}

\end{appendix}

\end{document}